\catcode`@=11
\global\newcount\secno
\global\newcount\subsecno
\global\newcount\subsubsecno
\global\newcount\paragraphno
\global\newcount\subparagraphno
\global\newcount\subsubparagraphno
\global\newcount\subsubsubparagraphno
\global\newcount\equationno
\global\newcount\refno
\global\newcount\footnoteno
\global\newcount\@no
\secno=0\subsecno=0\subsubsecno=0\equationno=0\refno=0\footnoteno=0
\global\newif\ifs@c \s@cfalse
\global\newif\ifsubs@c \subs@cfalse
\global\newif\ifsubsubs@c \subsubs@cfalse
\global\newif\ifp@ragraph \p@ragraphfalse
\global\newif\ifsubp@ragraph \subp@ragraphfalse
\global\newif\ifsubsubp@ragraph \subsubp@ragraphfalse
\global\newif\ifsubsubsubp@ragraph \subsubsubp@ragraphfalse
\newskip\normalsmallskipamount \normalsmallskipamount=3pt plus 1pt minus 1pt
\newskip\normalmedskipamount \normalmedskipamount=6pt plus 2pt minus 2pt
\newskip\normalbigskipamount \normalbigskipamount=12pt plus 4pt minus 4pt

\def\setbaselinestretch#1{\baselineskip=#1\normalbaselineskip
  \lineskip=#1\normallineskip
  \lineskiplimit=#1\normallineskiplimit
  \smallskipamount=#1\normalsmallskipamount
  \medskipamount=#1\normalmedskipamount
  \bigskipamount=#1\normalbigskipamount}
\let\\=\cr
\def\@draftleft#1{}
\def\@draftright#1{}
\overfullrule=0pt
\def\draft{\def\@draftleft##1{\leavevmode\vadjust{\smash{%
  \raise3pt\llap{\eighttt\string##1~~}}}}%
  \def\@draftright##1{\rlap{\eighttt~~\string##1}}%
  \def\date##1{\leftline{\number\month/\number\day/\number\year\
    \the\time}\bigskip}\overfullrule=5pt\setbaselinestretch{1.5}}
\def\@the#1{\ifnum\the#1>0\relax\the#1\else\ifnum\the#1<0\relax
  \@no=-\the#1\advance\@no'100\char\@no\else0\fi\fi}
\def\@advance#1{\ifnum\the#1<0\global\advance#1 -1\relax
  \else\global\advance#1 1\relax\fi}
\def\thesecno{\ifs@c\@the\secno\fi}
\def\thes@cno{\ifs@c\@the\secno.\fi}
\def\nsec#1\par{\bigbreak\bigskip\@advance\secno
  \subsecno=0\subsubsecno=0\equationno=0
  \paragraphno=0\subparagraphno=0\subsubparagraphno=0
  \subsubsubparagraphno=0
  \s@ctrue\subs@cfalse\subsubs@cfalse
  \p@ragraphfalse\subp@ragraphfalse\subsubp@ragraphfalse
  \subsubsubp@ragraphfalse
  \vbox{\secfont\noindent
  \thesecno. #1\medskip}\nobreak\noindent\ignorespaces}
\def\secadvance{\@advance\secno}
\def\sec#1#2\par{\bigbreak\bigskip
  \subsecno=0\subsubsecno=0\equationno=0
  \paragraphno=0\subparagraphno=0\subsubparagraphno=0
  \subsubsubparagraphno=0
  \s@ctrue\subs@cfalse\subsubs@cfalse
  \p@ragraphfalse\subp@ragraphfalse\subsubp@ragraphfalse
  \subsubsubp@ragraphfalse
  \if*#1\vbox{\secfont\noindent\ignorespaces#2\medskip}%
  \else
  \subsecno=0\subsubsecno=0\equationno=0
  \paragraphno=0\subparagraphno=0\subsubparagraphno=0
  \subsubsubparagraphno=0
  \s@ctrue\subs@cfalse\subsubs@cfalse
  \p@ragraphfalse\subp@ragraphfalse\subsubp@ragraphfalse
  \subsubsubp@ragraphfalse
  \secno=#1
  \vbox{\secfont\noindent\thesecno. #2\medskip}\fi
  \nobreak\noindent\ignorespaces}
\def\seclab#1{\xdef#1{\@the\secno}\@draftleft#1}
\def\thesubsecno{\thes@cno\ifsubs@c\@the\subsecno\fi}
\def\thesubs@cno{\thes@cno\ifsubs@c\@the\subsecno.\fi}
\def\nsubsec#1\par{\bigskip\@advance\subsecno
  \subsubsecno=0\equationno=0
  \paragraphno=0\subparagraphno=0\subsubparagraphno=0
  \subsubsubparagraphno=0
  \subs@ctrue\subsubs@cfalse
  \p@ragraphfalse\subp@ragraphfalse\subsubp@ragraphfalse
  \subsubsubp@ragraphfalse
  \vbox{\subsecfont\noindent\thesubsecno. #1\medskip}%
  \nobreak\noindent\ignorespaces}
\def\subsecadvance{\@advance\subsecno}
\def\subsec#1#2\par{\bigskip
  \subsubsecno=0\equationno=0
  \if*#1\vbox{\subsecfont\noindent\ignorespaces#2\medskip}%
  \else
  \subsubsecno=0\equationno=0
  \paragraphno=0\subparagraphno=0\subsubparagraphno=0
  \subsubsubparagraphno=0
  \subs@ctrue\subsubs@cfalse
  \p@ragraphfalse\subp@ragraphfalse\subsubp@ragraphfalse
  \subsubsubp@ragraphfalse
  \subsecno=#1
  \vbox{\subsecfont\noindent\thesubsecno. #2\medskip}\fi
  \nobreak\noindent\ignorespaces}
\def\subseclab#1{\xdef#1{\thesubsecno}%
  \@draftleft#1}
\def\thesubsubsecno{\thesubs@cno\ifsubsubs@c\@the\subsubsecno\fi}
\def\thesubsubs@cno{\thesubs@cno\ifsubsubs@c\@the\subsubsecno.\fi}
\def\nsubsubsec#1\par{\medskip\@advance\subsubsecno
  \paragraphno=0\subparagraphno=0\subsubparagraphno=0
  \subsubsubparagraphno=0
  \subsubs@ctrue
  \p@ragraphfalse\subp@ragraphfalse\subsubp@ragraphfalse
  \subsubsubp@ragraphfalse
  \vbox{\subsubsecfont\noindent
  \thesubsubsecno. #1\medskip}%
  \nobreak\noindent\ignorespaces}
\def\subsubsecadvance{\@advance\subsubsecno}
\def\subsubsec#1#2\par{\medskip
  \if*#1\vbox{\subsubsecfont\noindent\ignorespaces#2\medskip}%
  \else
  \paragraphno=0\subparagraphno=0\subsubparagraphno=0
  \subsubsubparagraphno=0
  \subsubs@ctrue
  \p@ragraphfalse\subp@ragraphfalse\subsubp@ragraphfalse
  \subsubsubp@ragraphfalse
  \subsubsecno=#1
  \vbox{\subsubsecfont\noindent
  \thesubsubsecno. #2\medskip}\fi
  \nobreak\noindent\ignorespaces}
\def\subsubseclab#1{\xdef#1{\thesubsubsecno}%
  \@draftleft#1}
\def\paragraph#1. {\par\thesubsubs@cno#1. \ignorespaces}
\def\theparagraphno{\thesubsubs@cno\ifp@ragraph\@the\paragraphno\fi}
\def\thep@ragraphno{\thesubsubs@cno\ifp@ragraph\@the\paragraphno.\fi}
\def\nparagraph{\@advance\paragraphno
  \subparagraphno=0\subsubparagraphno=0\subsubsubparagraphno=0
  \p@ragraphtrue\subp@ragraphfalse\subsubp@ragraphfalse
  \subsubsubp@ragraphfalse
  \par\theparagraphno. \ignorespaces}
\def\paragraphlab#1{\xdef#1{\theparagraphno}\@draftleft#1}
\def\thesubparagraphno{\thep@ragraphno\ifsubp@ragraph\@the\subparagraphno\fi}
\def\thesubp@ragraphno{\thep@ragraphno\ifsubp@ragraph\@the\subparagraphno.\fi}
\def\nsubparagraph{\@advance\subparagraphno
  \subsubparagraphno=0\subsubsubparagraphno=0
  \subp@ragraphtrue\subsubp@ragraphfalse
  \subsubsubp@ragraphfalse
  \par\thesubparagraphno. \ignorespaces}
\def\subparagraphlab#1{\xdef#1{\thesubparagraphno}\@draftleft#1}
\def\thesubsubparagraphno{\thesubp@ragraphno\ifsubsubp@ragraph
  \@the\subsubparagraphno\fi}
\def\thesubsubp@ragraphno{\thesubp@ragraphno\ifsubsubp@ragraph
  \@the\subsubparagraphno.\fi}
\def\nsubsubparagraph{\@advance\subsubparagraphno
  \subsubsubparagraphno=0
  \subsubp@ragraphtrue\subsubsubp@ragraphfalse
  \par\thesubsubparagraphno. \ignorespaces}
\def\subsubparagraphlab#1{\xdef#1{\thesubsubparagraphno}\@draftleft#1}
\def\thesubsubsubparagraphno{\thesubsubp@ragraphno\ifsubsubsubp@ragraph
  \@the\subsubsubparagraphno\fi}
\def\thesubsubsubp@ragraphno{\thesubsubp@ragraphno\ifsubsubsubp@ragraph
  \@the\subsubsubparagraphno.\fi}
\def\nsubsubsubparagraph{\@advance\subsubsubparagraphno
  \subsubsubp@ragraphtrue
  \par\thesubsubsubparagraphno. \ignorespaces}
\def\subsubsubparagraphlab#1{\xdef#1{\thesubsubsubparagraphno}\@draftleft#1}
\def\theequationno{\thesubs@cno\the\equationno}
\def\eqlabel#1{\global\advance\equationno1\xdef#1{\theequationno}%
  \eqno({\eqnofont #1})\@draftright#1}
\def\lnlabel#1{\global\advance\equationno1\xdef#1{\theequationno}%
  &({\eqnofont #1})\@draftright#1}
\def\eqadvance#1{\global\advance\equationno1\xdef#1{\theequationno}}
\def\eqlabelno(#1#2){\eqno({\eqnofont #1#2})\@draftright#1}
\def\lnlabelno(#1#2){&({\eqnofont #1#2})\@draftright#1}
\def\skips@c#1.{}
\def\skipsubs@c#1.#2.{}
\def\skipsubsubs@c#1.#2.{}
\def\skipsec#1{{\expandafter\skips@c#1}}
\def\skipsubsec#1{{\expandafter\skipsubs@c#1}}
\def\skipsubsec#1{{\expandafter\skipsubsubs@c#1}}
\newwrite\rfile
\def\nref#1#2{\global\advance\refno1\xdef#1{\the\refno}%
  \immediate\write
  \rfile{\noexpand\item{#1.}\noexpand\@draftleft\noexpand#1%
  #2}}
\def\sref#1#2{\immediate\write
  \rfile{\noexpand\item{#1.}\noexpand\@draftleft\noexpand#1%
  #2}}
\def\refs#1{\def\t@mp{#1}\futurelet\n@xtchar\t@stchar}
\def\t@stchar{\let\ignor@=\ign@re
  \ifx\n@xtchar..$^{\rm\t@mp}$\spacefactor=\sfcode`.{}%
  \else\ifx\n@xtchar,,$^{\rm\t@mp}$\spacefactor=\sfcode`,{}%
  \else\ifx\n@xtchar;;$^{\rm\t@mp}$\spacefactor=\sfcode`;{}%
  \else\ifx\n@xtchar::$^{\rm\t@mp}$\spacefactor=\sfcode`:{}%
  \else\ifx\n@xtchar??$^{\rm\t@mp}$\spacefactor=\sfcode`?{}%
  \else\ifx\n@xtchar!!$^{\rm\t@mp}$\spacefactor=\sfcode`!{}%
  \else$^{\rm\t@mp}$\let\ignor@=\v@id\fi\fi\fi\fi\fi\fi\ignor@}
\def\ign@re#1{}
\def\v@id{}
\def\Refs#1{$\rm#1$}
\def\reportno#1{\line{\hfil\vbox{\halign{\strut##\hfil\cr#1\crcr}}}}
\def\Title#1{\vskip3\bigskipamount\line{\titlefont
  \hfil\vbox{\halign{\strut\hfil##\hfil\cr#1\crcr}}\hfil}%
  \vskip2\bigskipamount}
\def\author#1{\centerline{\authorfont#1}\medskip}
\def\address#1{\centerline{\vbox{\halign
  {\strut\hfil\addressfont##\hfil\cr#1\crcr}}}%
  \bigskip}
\def\abstract#1{{\narrower\abstractfont\null\bigskip\noindent\ignorespaces
  #1\bigskip}}
\def\date#1{\leftline{#1}\bigskip}
\def\Tr{\mathop{\rm Tr}\nolimits}
\immediate\openout\rfile=refs.tmp
\font\seventeenrm=cmr17 \font\fourteenrm=cmr10 scaled 1440
\font\twelverm=cmr12  \font\eightrm=cmr8  \font\sixrm=cmr6
\font\seventeeni=cmmi10 scaled 1728 \font\fourteeni=cmmi10 scaled 1440
\font\twelvei=cmmi12  \font\eighti=cmmi8  \font\sixi=cmmi6
\font\seventeensy=cmsy10 scaled 1728 \font\fourteensy=cmsy10 scaled 1440
\font\twelvesy=cmsy10 scaled 1200 \font\eightsy=cmsy8 \font\sixsy=cmsy6
\font\seventeenbf=cmbx10 scaled 1728 \font\fourteenbf=cmbx10 scaled 1440
\font\twelvebf=cmbx12 \font\eightbf=cmbx8 \font\sixbf=cmbx6
\font\seventeentt=cmtt10 scaled 1728 \font\fourteentt=cmtt10 scaled 1440
\font\twelvett=cmtt12 \font\eighttt=cmtt8
\font\seventeenit=cmti10 scaled 1728 \font\fourteenit=cmti10 scaled 1440
\font\twelveit=cmti12 \font\eightit=cmti8
\font\seventeensl=cmsl10 scaled 1728 \font\fourteensl=cmsl10 scaled 1440
\font\twelvesl=cmsl12 \font\eightsl=cmsl8
\font\seventeenex=cmex10 scaled 1728 \font\fourteenex=cmex10 scaled 1440
\font\twelveex=cmex10 scaled 1200
\def\tenpoint{\def\rm{\fam0\tenrm}%
  \textfont0=\tenrm\scriptfont0=\sevenrm\scriptscriptfont0=\fiverm
  \textfont1=\teni\scriptfont1=\seveni\scriptscriptfont1=\fivei
  \textfont2=\tensy\scriptfont2=\sevensy\scriptscriptfont2=\fivesy
  \textfont3=\tenex\scriptfont3=\tenex\scriptscriptfont3=\tenex
  \textfont\itfam=\tenit\def\it{\fam\itfam\tenit}%
  \textfont\slfam=\tensl\def\sl{\fam\slfam\tensl}%
  \textfont\ttfam=\tentt\def\tt{\fam\ttfam\tentt}%
  \textfont\bffam=\tenbf\scriptfont\bffam=\sevenbf
  \scriptscriptfont\bffam=\fivebf\def\bf{\fam\bffam\tenbf}%
  \tenn@w
\def\small{\eightpoint}%
\def\large{\twelvepoint}%
\def\normalbaselines{\lineskip\normallineskip
  \baselineskip\normalbaselineskip \lineskiplimit\normallineskiplimit}%
  \setbox\strutbox=\hbox{\vrule height8.5pt depth3.5pt width0pt}%
  \normalbaselines\rm}
\def\twelvepoint{\def\rm{\fam0\twelverm}%
  \textfont0=\twelverm\scriptfont0=\eightrm\scriptscriptfont0=\sixrm
  \textfont1=\twelvei\scriptfont1=\eighti\scriptscriptfont1=\sixi
  \textfont2=\twelvesy\scriptfont2=\eightsy\scriptscriptfont2=\sixsy
  \textfont3=\twelveex\scriptfont3=\twelveex\scriptscriptfont3=\twelveex
  \textfont\itfam=\twelveit\def\it{\fam\itfam\twelveit}%
  \textfont\slfam=\twelvesl\def\sl{\fam\slfam\twelvesl}%
  \textfont\ttfam=\twelvett\def\tt{\fam\ttfam\twelvett}%
  \textfont\bffam=\twelvebf\scriptfont\bffam=\eightbf
  \scriptscriptfont\bffam=\sixbf\def\bf{\fam\bffam\twelvebf}%
  \twelven@w
  \def\small{\tenpoint}%
  \def\large{\fourteenpoint}%
  \def\normalbaselines{\lineskip1.2\normallineskip
    \baselineskip1.2\normalbaselineskip \lineskiplimit1.2\normallineskiplimit}%
  \setbox\strutbox=\hbox{\vrule height10.2pt depth4.2pt width0pt}%
  \normalbaselines\rm}
\def\fourteenpoint{\def\rm{\fam0\fourteenrm}%
  \textfont0=\fourteenrm\scriptfont0=\tenrm\scriptscriptfont0=\sevenrm
  \textfont1=\fourteeni\scriptfont1=\teni\scriptscriptfont1=\seveni
  \textfont2=\fourteensy\scriptfont2=\tensy\scriptscriptfont2=\sevensy
  \textfont3=\fourteenex\scriptfont3=\fourteenex\scriptscriptfont3=\fourteenex
  \textfont\itfam=\fourteenit\def\it{\fam\itfam\fourteenit}%
  \textfont\slfam=\fourteensl\def\sl{\fam\slfam\fourteensl}%
  \textfont\ttfam=\fourteentt\def\tt{\fam\ttfam\fourteentt}%
  \textfont\bffam=\fourteenbf\scriptfont\bffam=\tenbf
  \scriptscriptfont\bffam=\fivebf\def\bf{\fam\bffam\fourteenbf}%
  \fourteenn@w
  \def\small{\twelvepoint}%
  \def\large{\seventeenpoint}%
  \def\normalbaselines{\lineskip1.44\normallineskip
    \baselineskip1.44\normalbaselineskip \lineskiplimit1.44\normallineskiplimit}%
  \setbox\strutbox=\hbox{\vrule height12.24pt depth5.04pt width0pt}%
  \normalbaselines\rm}
\def\seventeenpoint{\def\rm{\fam0\seventeenrm}%
  \textfont0=\seventeenrm\scriptfont0=\twelverm\scriptscriptfont0=\eightrm
  \textfont1=\seventeeni\scriptfont1=\twelvei\scriptscriptfont1=\eighti
  \textfont2=\seventeensy\scriptfont2=\twelvesy\scriptscriptfont2=\eightsy
  \textfont3=\seventeenex\scriptfont3=\seventeenex
  \scriptscriptfont3=\seventeenex
  \textfont\itfam=\seventeenit\def\it{\fam\itfam\seventeenit}%
  \textfont\slfam=\seventeensl\def\sl{\fam\slfam\seventeensl}%
  \textfont\ttfam=\seventeentt\def\tt{\fam\ttfam\seventeentt}%
  \textfont\bffam=\seventeenbf\scriptfont\bffam=\twelvebf
  \scriptscriptfont\bffam=\eightbf
  \def\bf{\fam\bffam\seventeenbf}%
  \seventeenn@w
  \def\small{\fourteenpoint}%
  \def\large{\seventeenpoint}%
  \def\normalbaselines{\lineskip1.73\normallineskip
    \baselineskip1.73\normalbaselineskip \lineskiplimit1.73\normallineskiplimit}%
  \setbox\strutbox=\hbox{\vrule height14.7pt depth6.0pt width0pt}%
  \normalbaselines\rm}
\def\eightpoint{\def\rm{\fam0\eightrm}%
  \textfont0=\eightrm\scriptfont0=\sixrm\scriptscriptfont0=\fiverm
  \textfont1=\eighti\scriptfont1=\sixi\scriptscriptfont1=\fivei
  \textfont2=\eightsy\scriptfont2=\sixsy\scriptscriptfont2=\fivesy
  \textfont3=\tenex\scriptfont3=\tenex\scriptscriptfont3=\tenex
  \textfont\itfam=\eightit\def\it{\fam\itfam\eightit}%
  \textfont\slfam=\eightsl\def\sl{\fam\slfam\eightsl}%
  \textfont\ttfam=\eighttt\def\tt{\fam\ttfam\eighttt}%
  \textfont\bffam=\eightbf\scriptfont\bffam=\sixbf
  \scriptscriptfont\bffam=\fivebf\def\bf{\fam\bffam\eightbf}%
  \eightn@w
  \def\small{\eightpoint}%
  \def\large{\tenpoint}%
  \def\normalbaselines{\lineskip.8\normallineskip
    \baselineskip.8\normalbaselineskip \lineskiplimit.8\normallineskiplimit}%
  \setbox\strutbox=\hbox{\vrule height7pt depth3pt width0pt}%
  \normalbaselines\rm}
\def\small{\eightpoint}
\def\large{\twelvepoint}
\def\tenn@w{}
\def\twelven@w{}
\def\fourteenn@w{}
\def\seventeenn@w{}
\def\eightn@w{}
\def\newfont#1#2{\expandafter\font\csname ten#1\endcsname=#2
  \expandafter\font\csname twelve#1\endcsname=#2 scaled 1200
  \expandafter\font\csname fourteen#1\endcsname=#2 scaled 1440
  \expandafter\font\csname seventeen#1\endcsname=#2 scaled 1728
  \expandafter\font\csname eight#1\endcsname=#2 scaled 833
  \expandafter\font\csname seven#1\endcsname=#2 scaled 694
  \expandafter\font\csname six#1\endcsname=#2 scaled 579
  \expandafter\font\csname five#1\endcsname=#2 scaled 482
  \n@wfam{#1}{\csname #1fam\endcsname}\tenpoint}
\def\newf@m{\alloc@8\fam\chardef\sixt@@n}
\def\n@wfam#1#2{\expandafter\newf@m#2
  \let\t@mp=\tenn@w
  \global\xdef\tenn@w{\t@mp
    \textfont#2=\csname ten#1\endcsname
    \scriptfont#2=\csname seven#1\endcsname
    \scriptscriptfont#2=\csname five#1\endcsname
    \expandafter\def\csname#1\endcsname{\fam#2\csname ten#1\endcsname}}
  \let\t@mp=\twelven@w
  \global\xdef\twelven@w{\t@mp
    \textfont#2=\csname twelve#1\endcsname
    \scriptfont#2=\csname eight#1\endcsname
    \scriptscriptfont#2=\csname six#1\endcsname
    \expandafter\def\csname#1\endcsname{\fam#2\csname twelve#1\endcsname}}
  \let\t@mp=\fourteenn@w
  \global\xdef\fourteenn@w{\t@mp
    \textfont#2=\csname fourteen#1\endcsname
    \scriptfont#2=\csname ten#1\endcsname
    \scriptscriptfont#2=\csname seven#1\endcsname
    \expandafter\def\csname#1\endcsname{\fam#2\csname fourteen#1\endcsname}}
  \let\t@mp=\seventeenn@w
  \global\xdef\seventeenn@w{\t@mp
    \textfont#2=\csname seventeen#1\endcsname
    \scriptfont#2=\csname twelve#1\endcsname
    \scriptscriptfont#2=\csname eight#1\endcsname
    \expandafter\def\csname#1\endcsname{\fam#2\csname seventeen#1\endcsname}}
  \let\t@mp=\eightn@w
  \global\xdef\eightn@w{\t@mp
    \textfont#2=\csname eight#1\endcsname
    \scriptfont#2=\csname six#1\endcsname
    \scriptscriptfont#2=\csname five#1\endcsname
    \expandafter\def\csname#1\endcsname{\fam#2\csname eight#1\endcsname}}}
\def\vfootnote#1{\insert\footins\bgroup
  \interlinepenalty\interfootnotelinepenalty
  \splittopskip\ht\strutbox 
  \splitmaxdepth\dp\strutbox \floatingpenalty\@MM
  \leftskip\z@skip \rightskip\z@skip \spaceskip\z@skip \xspaceskip\z@skip
  \footnotefont\textindent{#1}\footstrut\futurelet\next\fo@t}
\def\nfootnote{\advance\footnoteno1\@no=\footnoteno\advance\@no'140
  \footnote{$^{\char\@no}$}}
\def\nvfootnote#1{\advance\footnoteno1\@no=\footnoteno\advance\@no'140
  \def#1{$^{\char\@no}$}\vfootnote{$^{\char\@no}$}}
%


%

\def\topcornerpagenumbers{\headline={\ifodd\the\pageno\hss\tenrm\folio\else
  \tenrm\folio\hss\fi}\footline={\hfil}}
%
\newdimen\columnsep \columnsep=0.1truein
\newdimen\fullhsize
\newbox\l@ftcolumn
\def\twocolumn{\supereject
  \fullhsize=\hsize
  \advance\hsize -\columnsep
  \divide\hsize 2
  \def\fullline{\hbox to\fullhsize}
  \let\column=L
  \output={\if L\column
      \global\setbox\l@ftcolumn=\c@lumnbox \global\let\column=R
    \else \doubleform@t \global\let\column=L\fi
    \ifnum\outputpenalty>-20000\else\dosupereject\fi}
  \def\doubleform@t{\shipout\vbox{\makeheadline
    \fullline{\box\l@ftcolumn\hfil\c@lumnbox}
    \makefootline}
  \advancepageno}
  \def\c@lumnbox{\leftline{\pagebody}}
  \def\makeheadline{\vbox to\z@{\vskip-22.5\p@
    \fullline{\vbox to8.5\p@{}\the\headline}\vss}\nointerlineskip}
  \def\makefootline{\baselineskip24\p@\fullline{\the\footline}}}
\def\endtwocolumn{\supereject
  \hsize=\fullhsize\output={\plainoutput}
  \def\makeheadline{\vbox to\z@{\vskip-22.5\p@
    \line{\vbox to8.5\p@{}\the\headline}\vss}\nointerlineskip}
  \def\makefootline{\baselineskip24\p@\line{\the\footline}}}
\def\:#1{\nobreak\hskip0pt\nobreak#1\nobreak\hskip0pt}
\def\eqalign#1{\null\,\vcenter{\openup\jot\m@th
  \ialign{\strut\hfil$\displaystyle{##}$&$\displaystyle{{}##}$\hfil
  &&\hfil$\displaystyle{##}$&$\displaystyle{{}##}$\hfil
      \crcr#1\crcr}}\,}
\def\titlefont{\seventeenpoint\rm}
\def\authorfont{\twelvepoint\rm}
\def\addressfont{\tenpoint\it}
\def\abstractfont{\eightpoint\rm}
\def\secfont{\tenpoint\bf}
\def\subsecfont{\tenpoint\sl}
\def\subsubsecfont{\tenpoint\it}
\def\eqnofont{\tenpoint\rm}
\def\footnotefont{\eightpoint\rm}
\catcode`@=12

\input pictex
\let\backslash=\\
\let\\=\cr

\def\refs#1{~[#1]}
\def\Refs#1{[#1]}

\def\Z{{\bf Z}}
\def\C{{\bf C}}
\def\F{{\cal F}}
\def\H{{\cal H}}
\def\O{{\cal O}}
\def\P{{\cal P}}
\def\Q{{\cal Q}}
\def\ve{\varepsilon}
\def\i{{\rm i}}
\def\e{{\rm e}}

\def\Im{\mathop{\rm Im}\nolimits}

\def\Tr{\mathop{\rm Tr}\nolimits}
\def\sh{\mathop{\rm sh}\nolimits}

\def\id{\mathop{\rm id}\nolimits}
\def\half{{1\over2}}
\def\quarter{{1\over4}}
\def\texthalf{{\textstyle{1\over2}}}

\def\overtext{\over\textstyle}
\def\lcolon{\mathopen:}
\def\rcolon{\mathclose:}

\def\[{\mathopen{[\![}}
\def\]{\mathclose{]\!]}}
\def\primephantom{{\vphantom{\prime}}}
\def\tf{{\tilde f}}
\def\tdt{{\tilde t}}
\def\tX{{\tilde X}}
\def\tY{{\tilde Y}}
\def\tV{{\tilde V}}
\def\bV{{\bar V}}
\def\tbV{{\tilde{\bar V}}}
\def\tR{{\tilde R}}
\def\W#1(#2,#3;#4,#5|#6){\mathop{W#1}\left[\matrix{#5&#4\cr#2&#3}\bigg|
                        \matrix{#6}\right]}
\def\tW#1(#2,#3;#4,#5|#6){\mathop{\tilde W#1}\left[\matrix{#5&#4\cr#2&#3}\bigg|
                        \matrix{#6}\right]}
\def\L(#1,#2;#3,#4|#5){\mathop L\left[\matrix{#4&#3\cr#1&#2}\bigg|
                        \matrix{#5}\right]}
\def\tL(#1,#2;#3,#4|#5){\mathop{\tilde L}\left[\matrix{#4&#3\cr#1&#2}\bigg|
                        \matrix{#5}\right]}
\def\K(#1,#2;#3,#4|#5){\mathop K\left[\matrix{#4&#3\cr#1&#2}\bigg|
                        \matrix{#5}\right]}
\def\ar#1 #2, #3 #4 {\arrow <0.3truecm> [0.1,0.3] from #1 #2 to #3 #4 }
\catcode`!=11
\def\ln#1 #2, #3 #4 {\!start(#1,#2)\!ljoin(#3,#4)}
\catcode`!=12
\def\rl#1 #2, #3 #4 {\putrule from #1 #2 to #3 #4 }
\def\trl#1 #2, #3 #4 {\linethickness .8pt
	\putrule from #1 #2 to #3 #4 \linethickness .4pt}
\def\point#1 #2 {\put{\hbox{\kern -1pt .}} [Bl] at #1 #2 }
\def\bfpoint#1 #2 {%
\put{\hbox{\kern -2pt \raise -2.3pt%
\hbox{$\textstyle\bullet$}}} [Bl] at #1 #2 }
\def\figcap#1#2\par{\medskip{\narrower\eightpoint\noindent
	Fig.~#1. \ignorespaces#2\par}\medskip}

\reportno{hep-th/0103144}
\Title{Free Field Construction for the Eight-Vertex Model:\\
Representation for Form Factors}
\author{Michael Lashkevich}
\address{Landau Institute for Theoretical Physics,\\
142432 Chernogolovka, Russia}

\abstract{The free field realization of the eight-vertex model is
extended to form factors. It is achieved by constructing off-diagonal
with respect to the ground state sectors matrix elements of the
$\Lambda$ operator which establishes a relation between corner transfer
matrices of the eight-vertex model and of the SOS model. As an example,
the two-particle form factor of the $\sigma^z$ operator is evaluated
explicitly.}

\date{March 2001}

\nref\LPone{M.~Lashkevich and Ya.~Pugai, {\it Nucl.\ Phys.}\ {\bf B516 [FS]}
	(1998) 623 [hep-th/9710099]}
\nref\LPtwo{M.~Lashkevich and Ya.~Pugai, {\it JETP Lett.}\ {\bf68} (1998) 257
	[hep-th/9805216]}
\nref\JMbook{M.~Jimbo and T.~Miwa, {\it Algebraic analysis of solvable lattice
	models}, CBMS Regional Conference Series in Mathematics {\bf85}
	AMS (1994)}
\nref\Baxterevtripple{R.~J.~Baxter, {\it Ann.\ Phys.}\ {\bf76} (1973) 1; 26; 48}
\nref\FIJKMY{O.~Foda, K.~Iohara, M.~Jimbo, R.~Kedem, T.~Miwa, and H.~Yan,
	{\it Prog.\ Theor.\ Phys.\ Suppl.}\ {\bf118} (1995)~1 [hep-th/9405058]}
\nref\LukPug{S.~Lukyanov and Ya.~Pugai, {\it Nucl.\ Phys.}\ {\bf B473 [FS]}
	(1996) 631 [hep-th/9602074]}
\nref\SuthFW{B.~Sutherland, {\it J.~Math.\ Phys.}\ {\bf11} (1970) 3183;
	C.~Fan and F.~Y.~Wu, {\it Phys.\ Rev.}\ {\bf B2} (1970) 723}
\nref\Baxcor{R.~J.~Baxter, {\it Phil.\ Trans.\ Royal Soc.\ London}
	{\bf289} (1978) 315}
\nref\Baxterevzero{R.~J.~Baxter, {\it Phys.\ Rev.\ Lett.}\ {\bf26} (1971) 832;
	{\it Ann.\ Phys.\ (N.Y.)\/} {\bf 70} (1972) 193}
\nref\JMNeightvertex{M.~Jimbo, T.~Miwa, and A.~Nakayashiki, {\it J.~Phys.}\
	{\bf A26} (1993) 2199}
\nref\JMOrsos{M.~Jimbo, T.~Miwa, and Y.~Ohta, {\it Int.~J.\ Mod.\ Phys.}\
	{\bf A8} (1993) 1457}
\nref\DFJMN{B.~Davies, O.~Foda, M.~Jimbo, T.~Miwa, and A.~Nakayashiki,
	{\it J.~Phys.}\ {\bf A26} (1993) 2199}
\nref\Smirnovbook{F.~A.~Smirnov, {\it Form factors in completely
	integrable models of quantum field theory},
	World Scientific, Singapore (1992)}
\nref\BK{R.~J.~Baxter and S.~B.~Kelland, {\it J.~Phys.}\ {\bf C7} (1974) L403}
\nref\JLMP{M.~Jimbo, M.~Lashkevich, T.~Miwa, and Ya.~Pugai,
	{\it Phys.\ Lett.}\ {\bf A229} (1997) 285 [hep-th/9607177]}
\nref\Hara{Y.~Hara, {\it Nucl.\ Phys.}\ {\bf B572 [FS]} (2000) 574
	[math-ph/9910046]}
\nref\KKMMNN{S.-J.~Kang, M.~Kashiwara, K.~Misra, T.~Miwa, T.~Nakashima,
	and A.~Nakayashiki, {\it Int.\ J.\ Mod.\ Phys.}\ {\bf A7}
	({\it Suppl.}~{\bf1A}) (1992) 449}
\nref\Frons{C.~Fr\o nsdal, {\it Publ.\ RIMS, Kyoto Univ.}\ {\bf33} (1997) 91
	[q-alg/9606020]; {\it Lett.\ Math.\ Phys.}\ {\bf40} (1997) 117
	[q-alg/9611028]}
\nref\JKOS{M.~Jimbo, H.~Konno, S.~Odake, and J.~Shiraishi,
	preprint q-alg/9712029}
\nref\KMT{N.~Kitanine, J.~M.~Maillet, and V.~Terras,
	{\it Nucl.\ Phys.}\ {\bf B554} (1999) 647 [math-ph/9807020];
	preprint math-ph/9907019}
\nref\IKMT{A.~G.~Izergin, N.~Kitanine, J.~M.~Maillet, and V.~Terras,
	preprint solv-int/9812021}
\nref\NPT{A.~Nakayashiki, S.~Pakuliak, and V.~Tarasov,
	preprint q-alg/9712002}

\nsec Introduction

This paper continues the research started in~\Refs{\LPone,\LPtwo},
where an approach to calculating local correlation functions of the
eight-vertex/XYZ model based on the vertex operator approach
(see~\Refs{\JMbook} and references therein) and the vertex-face
correspondence\refs{\Baxterevtripple} was proposed. Here the method is
extended to provide a representation for form factors. As an example,
explicit expressions for the two-particle form factor of the $\sigma^z$
operator have been found.

Form factors are matrix elements of local operators in the basis of
eigenvectors of the transfer matrix or Hamiltonian. In principle, all
physical properties of a system can be expressed in terms of form
factors. In practice, form factors provide series for correlation
functions well converging off the critical point and finite expressions
for some other quantities of physical importance.

In the case of the six-vertex model the integral representation for
correlation functions and form factors is based on the
infinite-dimensional symmetry of the model on the infinite lattice,
described by the quantum affine algebra $U_q(\widehat{sl}_2)$, and on
the free field representation for the elements of the algebra and the
intertwining operators of its infinite-dimensional representations. The
free boson field takes its origin in the Cartan-type elements of the
algebra in the Drinfeld realization. In the case of the eight-vertex
model the Drinfeld realization of its symmetry algebra,
${\cal A}_{q,p}(\widehat{sl}_2)$\refs{\FIJKMY}, is unknown.

It was shown\refs{\LPone} that the correlation functions of the
eight-vertex model can be related to those of the solid-on-solid (SOS)
model with an insertion of a particular non-local operator $\Lambda$.
As the free field representation for the SOS model was
known\refs{\LukPug} it remained to find the free field realization of
the $\Lambda$ operator. Such a realization was found for the particular
matrix elements diagonal in the SOS ground state sector. Though that
was enough for correlation functions, the operators related to
excitations in the eight-vertex model being mapped onto the SOS model
contain non-zero elements off-diagonal in ground state sector. So the
construction for form factors remained unsolved problem. It is solved
here.

The structure of the paper is the following. In Sec.~2 the definitions
of the models are reminded and main notations are introduced. In Sec.~3
the vertex operator algebra that unifies the eight-vertex and SOS
models in the corner transfer matrix representation is described. In
Sec.~4 the free field realization of the $\Lambda$ operator is proposed
which is the main result of the paper. The simplest form factor is
computed in Sec.~5. Details of some calculations and some formulas
used in the paper are gathered in Appendices.

\nsec Basic models and vertex-face correspondence

Recall that the eight-vertex model\refs{\SuthFW} is a model of
classical statistical mechanics on a square lattice, where the
fluctuating variables (`spins') $\ve=\pm\equiv\pm1$ are associated to
links of the lattice, while the interaction is associated to vertices.
Namely, a Boltzmann weight $R^{\ve_3\ve_4}_{\ve_1\ve_2}$ is associated
to each vertex of the model (Fig.~1a). The total statistical weight of
the configuration of spin variables is proportional to the product of
local Boltzmann weights over the lattice. As a Boltzmann weight
$R^{\ve_3\ve_4}_{\ve_1\ve_2}$ is assumed to be nonzero for
$\ve_1+\ve_2+\ve_3+\ve_4=0\pmod4$ only, there are eight admissible
configurations around each vertex, which justifies the name of the
model. If, besides, the weights are invariant under simultaneous change
of all signs of spins, the model is integrable\refs{\Baxterevzero} and
this is the case under consideration. The nonzero Boltzmann weights are
$$
a=R^{++}_{++}=R^{--}_{--},
\quad
b=R^{+-}_{+-}=R^{-+}_{-+},
\quad
c=R^{-+}_{+-}=R^{+-}_{-+},
\quad
d=R^{--}_{++}=R^{++}_{--}.
$$
The $R$ matrix can be parametrized in elliptic functions. Namely,
consider it as a function $R(u)\equiv R(u;\epsilon,r)$ of the spectral
parameter $u$ and two additional parameters $\epsilon$ and $r$ with
matrix elements
$$
\eqalign{
a(u;\epsilon,r)
&=\rho(u;\epsilon,r)s(1-u;\epsilon,r),
\cr
b(u;\epsilon,r)
&=\rho(u;\epsilon,r)s(u;\epsilon,r),
\cr
c(u;\epsilon,r)
&=\rho(u;\epsilon,r)s(1;\epsilon,r),
\cr
d(u;\epsilon,r)
&=\rho(u;\epsilon,r)s(1-u;\epsilon,r)s(u;\epsilon,r)s(1;\epsilon,r),
}\eqlabel\EQRmatrix
$$
where
$$
s(u;\epsilon,r)={\theta_1\!\left({u\over2r};{\i\pi\over2\epsilon r}\right)
 \over\theta_2\!\left({u\over2r};{\i\pi\over2\epsilon r}\right)}.
$$
Here $\theta_i(u;\tau)$ ($i=1,\ldots,4$) are the Jacobi theta functions
with the quasiperiods $1$ and $\tau$ ($\Im\tau>0$).

We shall also use the multiplicative variables
$$
x=\e^{-\epsilon},
\qquad
p=x^{2r},
\qquad
z=x^{2u}
\eqlabel\EQmultvariables
$$
on equal grounds with the additive ones $\epsilon$, $r$, $u$.

\topinsert
%
%
\line{\hfil
\beginpicture
\setcoordinatesystem units <1cm,1cm> point at 0 -0.5
\rl 2.5 2 , 2.5 0.8 \ar 2.5 0.81 , 2.5 0.8
\rl 3 1.5 , 1.8 1.5 \ar 1.81 1.5 , 1.8 1.5
\put{$R(u-v)_{\ve_1\ve_2}^{\ve_3\ve_4}=$} [Br] at 1.2 1.4
\put{$\ve_1$} [Bl] at 2.6 0.9
\put{$\ve_2$} [Bl] at 1.9 1.7
\put{$\ve_3$} [Bl] at 2.6 1.9
\put{$\ve_4$} [Bl] at 3.0 1.6
\put{$u$} [t] at 2.5 0.6
\put{$v$} [rB] at 1.7 1.5
\put{$(a)$} [Bl]  at 1.6 -0.7
\setcoordinatesystem units <1cm,1cm> point at -4 0
\rl 1.0 3.3 , 1.0 0.7
\rl 2.0 3.3 , 2.0 0.7
\rl 3.0 3.3 , 3.0 0.7
\rl 0.7 3.0 , 3.3 3.0
\rl 0.7 2.0 , 3.3 2.0
\rl 0.7 1.0 , 3.3 1.0
\put{$+$} [lb] at 1.1 3.4
\put{$+$} [lb] at 0.4 3.1
\put{$+$} [lb] at 3.1 3.4
\put{$+$} [lb] at 2.4 3.1
\put{$+$} [lb] at 2.1 2.4
\put{$+$} [lb] at 1.4 2.1
\put{$+$} [lb] at 1.1 1.4
\put{$+$} [lb] at 0.4 1.1
\put{$+$} [lb] at 3.4 2.1
\put{$+$} [lb] at 3.1 1.4
\put{$+$} [lb] at 2.4 1.1
\put{$+$} [lb] at 2.1 0.4
\put{$-$} [lb] at 2.1 3.4
\put{$-$} [lb] at 1.4 3.1
\put{$-$} [lb] at 1.1 2.4
\put{$-$} [lb] at 0.4 2.1
\put{$-$} [lb] at 3.4 3.1
\put{$-$} [lb] at 3.1 2.4
\put{$-$} [lb] at 2.4 2.1
\put{$-$} [lb] at 2.1 1.4
\put{$-$} [lb] at 1.4 1.1
\put{$-$} [lb] at 1.1 0.4
\put{$-$} [lb] at 3.4 1.1
\put{$-$} [lb] at 3.1 0.4
\put{$(b)$} [B] at 4.0 -0.2
\setcoordinatesystem units <1cm,1cm> point at -8 0
\rl 1.0 3.3 , 1.0 0.7
\rl 2.0 3.3 , 2.0 0.7
\rl 3.0 3.3 , 3.0 0.7
\rl 0.7 3.0 , 3.3 3.0
\rl 0.7 2.0 , 3.3 2.0
\rl 0.7 1.0 , 3.3 1.0
\put{$-$} [lb] at 1.1 3.4
\put{$-$} [lb] at 0.4 3.1
\put{$-$} [lb] at 3.1 3.4
\put{$-$} [lb] at 2.4 3.1
\put{$-$} [lb] at 2.1 2.4
\put{$-$} [lb] at 1.4 2.1
\put{$-$} [lb] at 1.1 1.4
\put{$-$} [lb] at 0.4 1.1
\put{$-$} [lb] at 3.4 2.1
\put{$-$} [lb] at 3.1 1.4
\put{$-$} [lb] at 2.4 1.1
\put{$-$} [lb] at 2.1 0.4
\put{$+$} [lb] at 2.1 3.4
\put{$+$} [lb] at 1.4 3.1
\put{$+$} [lb] at 1.1 2.4
\put{$+$} [lb] at 0.4 2.1
\put{$+$} [lb] at 3.4 3.1
\put{$+$} [lb] at 3.1 2.4
\put{$+$} [lb] at 2.4 2.1
\put{$+$} [lb] at 2.1 1.4
\put{$+$} [lb] at 1.4 1.1
\put{$+$} [lb] at 1.1 0.4
\put{$+$} [lb] at 3.4 1.1
\put{$+$} [lb] at 3.1 0.4
\endpicture
\hfil}
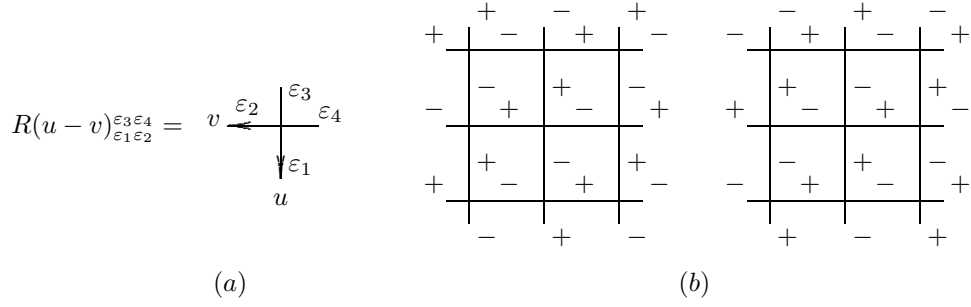
\figcap{1}Eight-vertex model: $(a)$ definition of the weight matrix;
$(b)$ two degenerate ground states.\par
\endinsert

The common factor $\rho(u;\epsilon,r)$ is convenient to be fixed so
that the partition function per site were equal to one. Under this
condition it reads
$$
\eqalign{
\rho(u;\epsilon,r)
&=x^{1-r/2}
{(x^{2r+2};x^{4r})_\infty(x^{2r-2};x^{4r})_\infty
\over(x^{2r};x^{4r})^2_\infty}
{(x^{2r}z;x^{4r})_\infty(x^{2r}z^{-1};x^{4r})_\infty
\over(x^{4r-2}z;x^{4r})_\infty(x^2z^{-1};x^{4r})_\infty}
{g(z^{-1})\over g(z)},
\cr
g(z)
&\equiv g(z;\epsilon,r)
={(x^2z;x^4,x^{2r})_\infty(x^{2r+2}z;x^4,x^{2r})_\infty
\over
(x^4z;x^4,x^{2r})_\infty(x^{2r}z;x^4,x^{2r})_\infty},
\cr
\span
(z;p_1,\ldots,p_N)_\infty
=\prod_{n_1,\ldots,n_N=0}^\infty(1-zp_1^{n_1}\ldots p_N^{n_N}).
}\eqlabel\EQrhofactor
$$

The matrix $R(u)$ satisfies the Yang--Baxter equation
$$
R_{12}(u_1-u_2)R_{13}(u_1-u_3)R_{23}(u_2-u_3)
=R_{23}(u_2-u_3)R_{13}(u_1-u_3)R_{12}(u_1-u_2),
$$
where $R_{ij}(u):V_i\otimes V_j\to V_i\otimes V_j$ acting on a tensor
product of several copies of identical spaces
$V_i\simeq\C^2$, and, with the common factor chosen in the form
(\EQrhofactor), the unitarity and crossing relations:
$$
R_{12}(u)R_{21}(-u)=\id,
\quad
R(u)_{\ve_1\ve_2}^{\ve_3\ve_4}
=R(1-u)_{\ve_4,-\ve_1}^{\ve_2,-\ve_3}.
$$

The region
$$
\epsilon>0,
\qquad
r>1,
\qquad
-1<u<1
$$
corresponds to the `antiferroelectric' region $a+|b|+|d|<c$. There are
two ground state configurations shown in Fig.~1b in this region. In
these ground states the $c$-type vertex configurations are only
realized. Any other configuration with finite energy differs from one
of these ground states by flopping a finite number of spins. We shall
speak about two ground state sectors in this model.

We shall consider the inhomogeneous generalization of the eight-vertex
model (the $Z$-invariant eight-vertex model\refs{\Baxcor}) in which a
value of the spectral parameter is associated to each line of the model
and the difference of spectral parameters at a vertex enters the
respective $R$ matrix (Fig.~1a).

Let us formulate now the solid-on-solid (SOS) model. This model was
proposed by Baxter\refs{\Baxterevtripple} as a partition function
preserving twist of the eight-vertex model that admits solution by
means of the Bethe Ansatz.

In the SOS model a fluctuating variable (`height') is associated to
each site of the lattice and runs over the set $\Z+\delta$ with some
$\delta$ satisfying $0<\delta<1$. The partition function is $\delta$
independent. We shall consider $\delta$ as a fixed parameter. The
heights $n_1$ and $n_2$ at adjacent sites satisfy the admissibility
condition
$$
|n_1-n_2|=1.
\eqlabel\EQadmissibility
$$
Interaction is described by the Boltzmann weights $\W(n_1,n_2;n_3,n_4|u)$
associated to the faces of the lattice, while the spectral parameters
are attached to the lines of the dual lattice (Fig.~2a):
$$
\eqalign{
\W(n\pm1,n\pm2;n\pm1,n|u)
&=R_0(u),
\cr
\W(n\pm1,n;n\pm1,n|u)
&=R_0(u){[n\pm u][1]\over [n][1-u]},
\cr
\W(n\mp1,n;n\pm1,n|u)
&=-R_0(u){[n\pm1][u]\over [n][1-u]}.
}\eqlabel\EQWmatrix
$$
Here
$$
R_0(u)\equiv R_0(u;\epsilon,r)=z^{(r-1)/2r}{g(z^{-1})\over g(z)}
\eqlabel\EQRofactor
$$
with $g(z)$ from (\EQrhofactor) and here and below we use the assignments
$$
\eqalign{
[u]_i
&=h_i(u;r)=\sqrt{\pi\over\epsilon r}\,\e^{\quarter\epsilon r}\,
\theta_i\!\left({u\over r};{\i\pi\over\epsilon r}\right),
\cr
[u]
&=[u]_1=x^{u^2/r-u}(z;p)_\infty(pz^{-1};p)_\infty(p;p)_\infty.
}\eqlabel\EQhfunctions
$$
For brevity, we omitted the variables $\epsilon$ and $r$ in the argument
of the $W$s, but we shall restore them when necessary. Generally, for any
function $f$ of these variables we shall imply $f(u)\equiv f(u;\epsilon,r)$.

\topinsert
%
%
\line{\hfil
\beginpicture
\setcoordinatesystem units <1cm,1cm> point at 0 0
\put{$\W(n_1,n_2;n_3,n_4|u-v)=$} [rB] at 3.7 2.4
\rl 5.0 2.0 , 6.0 2.0
\rl 6.0 2.0 , 6.0 3.0
\rl 6.0 3.0 , 5.0 3.0
\rl 5.0 3.0 , 5.0 2.0
\setdashes <2.5pt>
\rl 5.5 3.3 , 5.5 1.5
\rl 6.3 2.5 , 4.5 2.5
\setsolid
\ar 5.5 1.51 , 5.5 1.5
\ar 4.51 2.5 , 4.5 2.5
\put{$n_1$} [tr] at 4.9 1.9
\put{$n_2$} [tl] at 6.1 1.9
\put{$n_3$} [bl] at 6.1 3.1
\put{$n_4$} [br] at 4.9 3.1
\put{$u$} [t]  at 5.5 1.4
\put{$v$} [rB] at 4.4 2.5
\put{$(a)$} [B] at 3.5 0.0
\setcoordinatesystem units <1cm,1cm> point at -7 0
\rl 1.0 1.0 , 3.0 1.0
\rl 1.0 2.0 , 3.0 2.0
\rl 1.0 3.0 , 3.0 3.0
\rl 1.0 1.0 , 1.0 3.0
\rl 2.0 1.0 , 2.0 3.0
\rl 3.0 1.0 , 3.0 3.0
\put{\small$m+1$} [lB] at 1.05 3.1
\put{\small$m+1$} [lB] at 3.05 3.1
\put{\small$m+1$} [lB] at 2.05 2.1
\put{\small$m+1$} [lB] at 1.05 1.1
\put{\small$m+1$} [lB] at 3.05 1.1
\put{\small$m$} [lB] at 2.05 3.1
\put{\small$m$} [lB] at 1.05 2.1
\put{\small$m$} [lB] at 3.05 2.1
\put{\small$m$} [lB] at 2.05 1.1
\rl 4.0 1.0 , 6.0 1.0
\rl 4.0 2.0 , 6.0 2.0
\rl 4.0 3.0 , 6.0 3.0
\rl 4.0 1.0 , 4.0 3.0
\rl 5.0 1.0 , 5.0 3.0
\rl 6.0 1.0 , 6.0 3.0
\put{\small$m$} [lB] at 4.05 3.1
\put{\small$m$} [lB] at 6.05 3.1
\put{\small$m$} [lB] at 5.05 2.1
\put{\small$m$} [lB] at 4.05 1.1
\put{\small$m$} [lB] at 6.05 1.1
\put{\small$m+1$} [lB] at 5.05 3.1
\put{\small$m+1$} [lB] at 4.05 2.1
\put{\small$m+1$} [lB] at 6.05 2.1
\put{\small$m+1$} [lB] at 5.05 1.1
\put{$(b)$} [B] at 3.5 0.0
\endpicture
\hfil}
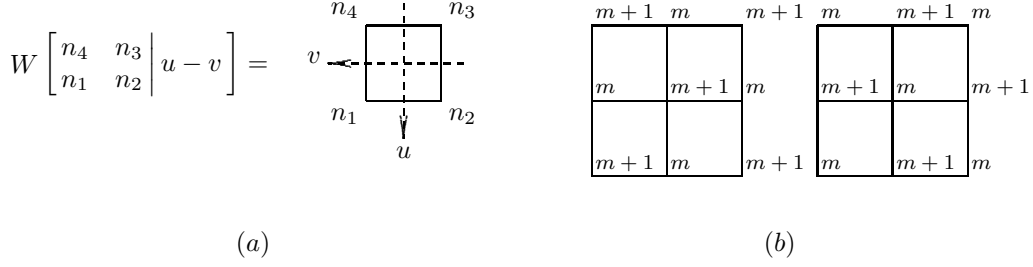
\figcap{2} SOS model: $(a)$ definition of weights; $(b)$ an infinite
number of degenerate ground states labeled by $m\in\Z+\delta$
subject to $(k-1)r<m,m+1<kr$ for some integer $k$ and by position of
the lesser value.\par
\endinsert

The weights $W(u)$ satisfy the dynamical Yang--Baxter equation and,
with the normalization (\EQRofactor), the dynamical versions of the
unitarity and crossing conditions. The advantage of the SOS model is
absence of a $d$-type configuration, which makes it possible to apply
the techniques developed for the six-vertex/XXZ model.

We are interested in the so called regime~III:
$$
\epsilon>0,\qquad r\geq1, \qquad 0<u<1.
\eqlabel\EQregimeIII
$$
There is an infinite number of degenerate, but inequivalent, ground states
in this regime shown in Fig.~2b. The ground state sectors will be labelled
by the value of $m$. The two pictures of Fig.~2b are distinguished by
parity of $m-n$ with $n$ being a value of the height at some given site.

Consider the functions (Fig.~3a)
$$
\eqalign{
t_+(u)^{n'}_n
&=\i^{1/2}f(u)
\theta_3\!\left({(n'-n)u+n'\over2r};\i{\pi\over2\epsilon r}\right),
\cr
t_-(u)^{n'}_n
&=\i^{-1/2}(-1)^{n-m_0+1}f(u)
\theta_4\!\left({(n'-n)u+n'\over2r};\i{\pi\over2\epsilon r}\right)
}\eqlabel\EQintertwining
$$
with $m_0$ being an arbitrary number from $\Z+\delta$ and $n$, $n'$
being an admissible pair of height variables. It is convenient to
choose the common factor $f(u)$ so that
$$
[u]f(u)f(u-1)=C\equiv{[0]_4^2\over
2\theta_3(0;\i\pi/2\epsilon r)\theta_4(0;\i\pi/2\epsilon r)}.
\eqlabel\EQCdef
$$
In~\Refs{\LPone} $f(u)$ was chosen in the form
$$
\eqalign{
f(u)
&=\sqrt{C}x^{-u^2/2r+(r-1)/2r+1/4} f_1(x^{2u}),
\cr
f_1(z)
&={1\over\sqrt{(x^{2r};x^{2r})_\infty}}
{(x^4z;x^4,x^{2r})_\infty(x^{2+2r}z^{-1};x^4,x^{2r})_\infty
\over(x^2z;x^4,x^{2r})_\infty(x^{2r}z^{-1};x^4,x^{2r})_\infty}.
}\eqlabel\EQffunction
$$

\topinsert
%
%
\line{\hfil
\beginpicture
\setcoordinatesystem units <1cm,1cm> point at 0 0
\put{$t_\ve(u_0-u)^{n'}_n=$} [rB] at 3.2 4.4
\trl 4.0 4.5 , 5.0 4.5
\ln 4.4 4.5 , 4.5 4.4
\ln 4.5 4.4 , 4.6 4.5
\rl 4.5 4.4 , 4.5 3.3 \ar 4.5 3.31 , 4.5 3.3
\setdashes <2.5pt>
\rl 3.8 4.2 , 5.2 4.2
\setsolid
\ar 3.81 4.2 , 3.8 4.2
\put{$n$} [b] at 5.0 4.6
\put{$n'$} [b] at 4.0 4.6
\put{$\ve$} [lB] at 4.6 3.8
\put{$u$} [t] at 4.5 3.2
\put{$u\smash{{}_0}$} [r] at 3.7 4.2
\put{$t^*_\ve(u_0-u)^{n'}_n=$} [rB] at 3.2 1.5
\trl 4.0 1.5 , 5.0 1.5
\ln 4.4 1.5 , 4.5 1.6
\ln 4.5 1.6 , 4.6 1.5
\rl 4.5 1.6 , 4.5 2.4
\setdashes <2.5pt>
\rl 4.5 0.9 , 4.5 1.5
\rl 3.8 1.8 , 5.2 1.8
\setsolid
\ar 4.5 0.91 , 4.5 0.9
\ar 3.81 1.8 , 3.8 1.8
\put{$n$} [rt] at 4.0 1.4
\put{$n\smash{{}'}$} [lt] at 5.0 1.4
\put{$\ve$} [lB] at 4.6 2.0
\put{$u$} [t] at 4.5 0.8
\put{$u\smash{{}_0}$} [r] at 3.7 1.8
\put{$(a)$} [B] at 3.7 0.0
\setcoordinatesystem units <1cm,1cm> point at -7 0
\trl 1.0 4.5 , 2.0 4.5
\ln 1.4 4.5 , 1.5 4.4 \ln 1.5 4.4 , 1.6 4.5
\trl 1.0 3.5 , 2.0 3.5
\ln 1.4 3.5 , 1.5 3.6 \ln 1.5 3.6 , 1.6 3.5
\rl 1.5 4.4 , 1.5 3.6
\setdots <2.5pt>
\rl 1.0 4.5 , 1.0 3.5
\setsolid
\put{$n$} [rB] at 0.9 4.6
\put{$n$} [rt] at 0.9 3.4
\put{$n\smash{{}'}$} [lt] at 2.1 3.4
\put{$n''$} [lB] at 2.1 4.6
\put{$t$}   [B] at 1.5 4.6
\put{$t^*$} [t] at 1.5 3.4
\put{$=\delta_{n'n''}$} [lB] at 2.9 4.0
\trl 1.0 2.5 , 2.0 2.5
\ln 1.4 2.5 , 1.5 2.4 \ln 1.5 2.4 , 1.6 2.5
\linethickness=1pt
\rl 1.41 2.49 , 1.59 2.49 \rl 1.44 2.46 , 1.56 2.46
\rl 1.43 2.47 , 1.57 2.47 \rl 1.47 2.43 , 1.53 2.43
\linethickness=.4pt
\trl 1.0 1.5 , 2.0 1.5
\ln 1.4 1.5 , 1.5 1.6 \ln 1.5 1.6 , 1.6 1.5
\rl 1.5 2.4 , 1.5 1.6
\setdots <2.5pt>
\rl 2.0 2.5 , 2.0 1.5
\setsolid
\put{$n''$} [rB] at 0.95 2.6
\put{$n\smash{{}'}$} [rt] at 0.95 1.4
\put{$n$} [lt] at 2.1 1.4
\put{$n$} [lB] at 2.1 2.6
\put{$t'$}  [B] at 1.5 2.6
\put{$t^*$} [t] at 1.5 1.4
\put{$=\delta_{n'n''}$} [lB] at 2.9 2.0
\put{$(b)$} [B] at 2.5 0.0
\endpicture
\hfil}
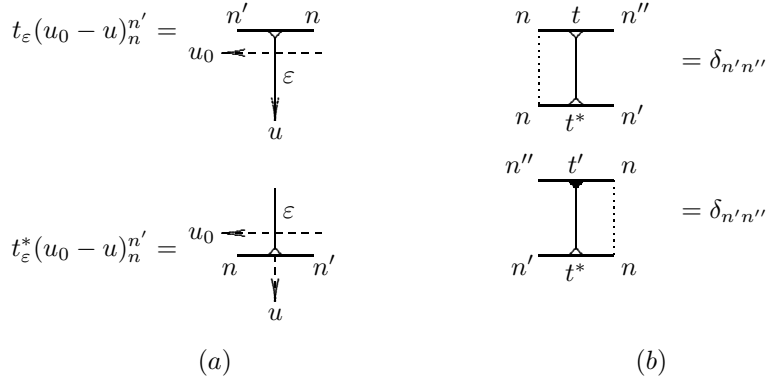
\figcap{3} Intertwining vectors: $(a)$ graphic notations; $(b)$ definition
of the conjugate and `primed' vectors.\par
\endinsert

The functions $t_\ve(u)^{n'}_n$ (called intertwining vectors) relate
the vertex and face type $R$ matrices (Fig.~4a):
$$
\sum_{\ve'_1\ve'_2}R(u-v)_{\ve_1\ve_2}^{\ve'_1\ve'_2}
t_{\ve'_1}(u_0-u)^{n'}_{s'} t_{\ve'_2}(u_0-v)^{s'}_n
=\sum_{s\in\Z} t_{\ve_2}(u_0-v)^{n'}_s t_{\ve_1}(u_0-u)^s_n
\W(s,n;s',n'|u-v)
\eqlabel\EQvertexface
$$
The last relation is referred to as the vertex-face correspondence.

\topinsert
%
%
\line{\hfil
\beginpicture
\setcoordinatesystem units <1cm,1cm> point at 0 0
\trl 1.0 2.5 , 2.0 2.5
\ln 1.4 2.5 , 1.5 2.4 \ln 1.5 2.4 , 1.6 2.5
\trl 2.0 2.5 , 2.0 1.5
\ln 2.0 2.1 , 1.9 2.0 \ln 1.9 2.0 , 2.0 1.9
\rl 1.5 2.4 , 1.5 1.1 \ar 1.5 1.11 , 1.5 1.1
\rl 1.9 2.0 , 0.6 2.0 \ar 0.61 2.0 , 0.6 2.0
\put{$n$}  [lt] at 2.1 1.4
\put{$s'$} [lB] at 2.1 2.6
\put{$n'$} [rB] at 0.9 2.6
\put{$\ve_1$} [lB] at 1.6 1.4
\put{$\ve_2$} [lt] at 1.0 1.9
\put{$u$} [t]  at 1.5 1.0
\put{$v$} [rB] at 0.5 1.95
\put{$=$} [B] at 2.75 1.95
\trl 4.0 2.5 , 4.0 1.5
\ln 4.0 2.1 , 3.9 2.0 \ln 3.9 2.0 , 4.0 1.9
\trl 4.0 1.5 , 5.0 1.5
\ln 4.4 1.5 , 4.5 1.4 \ln 4.5 1.4 , 4.6 1.5
\rl 4.0 2.5 , 5.0 2.5
\rl 5.0 2.5 , 5.0 1.5
\rl 3.9 2.0 , 3.4 2.0 \ar 3.41 2.0 , 3.4 2.0
\rl 4.5 1.4 , 4.5 0.9 \ar 4.5 0.91 , 4.5 0.9
\setdashes <2.5pt>
\rl 5.3 2.0 , 4.0 2.0
\rl 4.5 2.8 , 4.5 1.5
\setsolid
\put{$n$}  [lt] at 5.1 1.4
\put{$s'$} [lB] at 5.1 2.6
\put{$n'$} [rB] at 3.9 2.6
\put{$\ve_1$} [lB] at 4.65 1.1
\put{$\ve_2$} [lt] at 3.6 1.85
\put{$u$} [t]  at 4.5 0.8
\put{$v$} [rB] at 3.3 1.95
\put{$(a)$} [B] at 3.0 0.0
\setcoordinatesystem units <1cm,1cm> point at -7 0
\trl 1.0 2.5 , 1.0 1.5
\ln 1.0 2.1 , 1.1 2.0 \ln 1.1 2.0 , 1.0 1.9
\trl 1.0 1.5 , 2.0 1.5
\ln 1.4 1.5 , 1.5 1.6 \ln 1.5 1.6 , 1.6 1.5
\rl 1.1 2.0 , 2.0 2.0
\rl 1.5 1.6 , 1.5 2.5
\setdashes <2.5pt>
\rl 1.0 2.0 , 0.5 2.0
\rl 1.5 1.5 , 1.5 1.0
\setsolid
\ar 0.51 2.0 , 0.5 2.0
\ar 1.5 1.01 , 1.5 1.0
\put{$n$}  [lt] at 2.1 1.4
\put{$s$}  [rt] at 0.9 1.4
\put{$n'$} [rB] at 1.0 2.6
\put{$\ve'_1$} [lB] at 1.6 2.5
\put{$\ve'_2$} [lB] at 2.0 2.1
\put{$u$} [t]  at 1.5 0.9
\put{$v$} [rB] at 0.4 1.95
\put{$=$} [B] at 2.75 1.95
\trl 5.0 1.5 , 5.0 2.5
\ln 5.0 1.9 , 5.1 2.0 \ln 5.1 2.0 , 5.0 2.1
\trl 5.0 2.5 , 4.0 2.5
\ln 4.6 2.5 , 4.5 2.6 \ln 4.5 2.6 , 4.4 2.5
\rl 5.0 1.5 , 4.0 1.5
\rl 4.0 1.5 , 4.0 2.5
\rl 5.1 2.0 , 5.3 2.0
\rl 4.5 2.6 , 4.5 2.8
\setdashes <2.5pt>
\rl 5.0 2.0 , 3.4 2.0
\rl 4.5 2.5 , 4.5 1.0
\setsolid
\ar 3.41 2.0 , 3.4 2.0
\ar 4.5 1.01 , 4.5 1.0
\put{$n$}  [lt] at 5.1 1.4
\put{$s$}  [rt] at 3.9 1.4
\put{$n'$} [rB] at 3.9 2.6
\put{$\ve'_1$} [lB] at 4.5 2.8
\put{$\ve'_2$} [lB] at 5.3 2.1
\put{$u$}  [t] at 4.5 0.9
\put{$v$} [rB] at 3.3 1.95
\put{$(b)$} [B] at 3.0 0.0
\endpicture
\hfil}
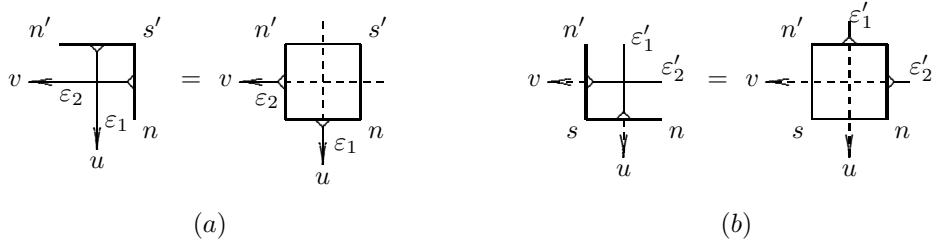
\figcap{4} Vertex-face correspondence: $(a)$~by usual intertwining vectors;
$(b)$~by conjugate intertwining vectors.\par
\endinsert

It is convenient to introduce the `conjugate' and `primed' intertwining
vectors (Fig.~3b):
$$
\sum_\ve t^*_\ve(u)^{n'}_n t_\ve(u)_{n''}^n = \delta_{n'n''},
\qquad
\sum_\ve t^*_\ve(u)^n_{n'} t'_\ve(u)_n^{n''} = \delta_{n'n''}.
\eqlabel\EQstarprimedef
$$
Explicitly, with $f(u)$ subjected to (\EQCdef) we have
$$
\eqalign{
t^*_\ve(u)_n^{n'}
&=(-)^{n-m_0+1}{(n'-n)\over [n]}\,t_{-\ve}(u-1)^{n'}_n,
\cr
t'_\ve(u)_n^{n'}
&={[n']\over [n]}\,t_\ve(u-2)_n^{n'}.
}\eqlabel\EQstarprimeexplicit
$$
The conjugate intertwining vector enters another form of the
vertex-face correspondence (Fig.~4b):
$$
\sum_{\ve_1\ve_2}
t^*_{\ve_2}(u_0-v)_{n'}^s t^*_{\ve_1}(u_0-u)_s^n
R(u-v)^{\ve'_1\ve'_2}_{\ve_1\ve_2}
=\sum_{s'\in\Z}
\W(s,n;s',n'|u-v)
t^*_{\ve'_1}(u_0-u)_{n'}^{s'} t^*_{\ve'_2}(u_0-v)_{s'}^n.
\eqlabel\EQvertexfaceconj
$$
For $t'_\ve(u)^{n'}_n$ the identity of the form (\EQvertexface) is valid.

\nsec Vertex operator algebra

Algebras of vertex operators describe models of statistical mechanics
in the corner transfer matrix approach (see~\Refs{\JMbook} and
references therein). Let us recall the structures of these algebras
for the eight-vertex and SOS models\refs{\JMNeightvertex,\JMOrsos}.

For a given orthogonal basis $|+\rangle$ and $|-\rangle$ in $\C^2$,
consider the subspace $\H_i$ in $\C^2\otimes\C^2\otimes\cdots$ spanned
on the vectors
$|\ve_1\rangle\otimes|\ve_2\rangle\otimes\cdots$
such that $\ve_k$ stabilizes to $(-)^{k+i}$ for large $k$. It means
that the number $i=0,1$ labels the ground state sector. We shall
identify this space with a right half row of vertical links on the
lattice. All operators are thought to act in the upper direction on
the right half lattice.

Define the operators
$$
\rho^{(i)}=x^{4H^{(i)}}:\H_i\to\H_i,
\qquad
\Phi^{(1-i,i)}_\ve(u),\Psi^{*(1-i,i)}_\ve(u):\H_i\to\H_{1-i}.
\eqlabel\EQevctmvos
$$
The spectrum of the corner Hamiltonian $H^{(i)}$ is equidistant and is
determined by the generating function
$$
\chi(q)\equiv\Tr_{\H_0}q^{H^{(0)}}=\Tr_{\H_1}q^{H^{(1)}}
=(q^{1/2};q)_\infty^{-1}.
\eqlabel\EQcharacter
$$
In the $H^{(i)}$-graded basis the matrix elements of the operators
$\Phi$ and $\Psi^*$ are meromorphic functions of the parameters
$z^{1/2}=x^u$ and $z=x^{2u}$, respectively, with the only pole at
$z^{1/2}=0$ ($z=0$).

The vertex operators $\Phi^{(1-i,i)}_\ve(u)$, $\Psi^{*(1-i,i)}_\ve(u)$
satisfy the commutation relations%
\nfootnote{Up to now this is not a theorem, but a well-checked
conjecture\refs{\DFJMN}.}
$$
\eqalign{
\span
\Phi^{(i,1-i)}_{\ve_1}(u_1)\Phi^{(1-i,i)}_{\ve_2}(u_2)
=\sum_{\ve'_1\ve'_2}R(u_1-u_2)^{\ve'_1\ve'_2}_{\ve_1\ve_2}
\Phi^{(i,1-i)}_{\ve'_2}(u_2)\Phi^{(1-i,i)}_{\ve'_1}(u_1),
\cr
\span
\sum_{\ve'_1\ve'_2}\tR(u_1-u_2)^{\ve'_1\ve'_2}_{\ve_1\ve_2}
\Psi^{*(i,1-i)}_{\ve'_1}(u_1)\Psi^{*(1-i,i)}_{\ve'_2}(u_2)
=\Psi^{*(i,1-i)}_{\ve_2}(u_2)\Psi^{*(1-i,i)}_{\ve_1}(u_1),
\cr
\span
\Phi^{(i,1-i)}_{\ve_1}(u_1)\Psi^{*(1-i,i)}_{\ve_2}(u_2)
=\tau(u_1-u_2)\Psi^{*(i,1-i)}_{\ve_2}(u_2)\Phi^{(1-i,i)}_{\ve_1}(u_1)
}\eqlabel\EQevcommutation
$$
and the operator $\rho^{(i)}$ shifts the spectral parameter:
$$
\Phi^{(1-i,i)}_\ve(u)\rho^{(i)}=\rho^{(1-i)}\Phi^{(1-i,i)}_\ve(u-2),
\qquad
\Psi^{*(1-i,i)}_\ve(u)\rho^{(i)}=\rho^{(1-i)}\Psi^{*(1-i,i)}_\ve(u-2).
\eqlabel\EQevdiff
$$
Here $\tR(u)$ is the $R$ matrix with the matrix elements
$$
\tilde a(u)=-a(u;\epsilon,r-1),
\quad
\tilde b(u)=-b(u;\epsilon,r-1),
\quad
\tilde c(u)=-c(u;\epsilon,r-1),
\quad
\tilde d(u)=d(u;\epsilon,r-1),
\eqlabel\EQtRdef
$$
and
$$
\tau(u)=\i{\theta_1\!\left(\quarter-{u\over2};{\i\pi\over2\epsilon}\right)
\over\theta_1\!\left(\quarter+{u\over2};{\i\pi\over2\epsilon}\right)}.
\eqlabel\EQtaufunction
$$

Besides, the vertex operators satisfy the additional relations
$$
\eqalign{
\span
\sum_\ve\Phi^{*(i,1-i)}_\ve(u)\Phi^{(1-i,i)}_\ve(u)=1,
\qquad
\Phi^{(i,1-i)}_{\ve_1}(u)\Phi^{*(1-i,i)}_{\ve_2}(u)=\delta_{\ve_1\ve_2},
\cr
\span
\Psi^{(i,1-i)}_{\ve_1}(u')\Psi^{*(1-i,i)}_{\ve_2}(u)
={1\over\pi}{\delta_{\ve_1\ve_2}\over u'-u}+O((u'-u)^0),
}\eqlabel\EQevnorm
$$
where
$$
\Phi^{*(1-i,i)}_\ve(u)=\Phi^{(1-i,i)}_{-\ve}(u-1),
\qquad
\Psi^{(1-i,i)}_\ve(u)=\Psi^{*(1-i,i)}_{-\ve}(u-1).
\eqlabel\EQevconj
$$
Below we will usually omit the superscripts indicating the ground state
sector and write simply $\rho$, $\Phi_\ve(u)$ {\it etc.}

The operator $\rho$ is nothing but the product of four corner
transfer matrices of the model. The operators $\Phi_\ve(u)$
(type~I vertex operators) form the `half transfer matrix' which is a
half-infinite version of the monodromy operator. Products of the
operators $\Psi_\ve(u)$ (type~II vertex operators) represent
many-particle eigenstates of the transfer matrix. The function $\tau(u)$
describes the excitation spectrum of the model while the function
$\tR(u)$ is the two-particle $S$ matrix of the elementary excitations.

Let $|a_1\theta_1,\ldots,a_N\theta_N\rangle^{(i)}$ be an eigenvector of
the transfer matrix corresponding to $N$ `charged' particles with
charges $a_j$ and `rapidities' $\theta_j=\pi^2/2\epsilon+\i\pi v_j$,
normalized by the condition
$$
{}^{(i)}\langle a'_1\theta'_1,\ldots,a'_N\theta'_N|
a_1\theta_1,\ldots,a_N\theta_N\rangle^{(i)}
=(2\pi)^N\sum_{\sigma\in S_N}\prod_{j=1}^N
\delta_{a'_{\sigma(j)}a_j}\delta(\theta'_{\sigma(j)}-\theta_j).
$$
`Rapidities' are implied to be real numbers and defined modulo
$\pi^2/\epsilon$. The vacuum state, which corresponds to $N=0$, will be
designated as $|0\rangle^{(i)}$. For $r<2$ this basis is incomplete
because of presence of `neutral' breather-like bound states.
Nevertheless, the form factors with these bound states are expressed in
terms of the form factors that only contain `charged' particles, though
with complex rapidities.

Let $\H_{i,i}$ be a subspace of the tensor product
$\cdots\otimes\C^2\otimes\C^2\otimes\cdots$ spanned on the vectors
$\cdots\otimes|\ve_1\rangle\otimes|\ve_2\rangle\otimes\cdots$ such that
$\ve_k=(-)^{k+i+1}$ for $k\to\infty$ and for
$k\to-\infty$. We shall identify this space with a column of horizontal
links in the lattice of the eight-vertex model. Let
$E_{\ve'\ve}:\C^2\to\C^2$ be a matrix with the only nonzero matrix
element $(\ve',\ve)$ equal to $1$. Respectively, the operator
$E^{(n)}_{\ve'\ve}:\H_{i,i}\to\H_{i,i}$ acts as $E_{\ve'\ve}$ on the
$n$th tensor component of $\H_{i,i}$ and as the identity mapping on
other components. To the operator
$$
\O=E^{(1)}_{\ve'_1\ve^\primephantom_1}\ldots
E^{(M)}_{\ve'_M\ve^\primephantom_M}
$$
in $\H_{i,i}$ we assign the operator
$$
\hat\O=\Phi^*_{\ve'_M}(u_M)\ldots\Phi^*_{\ve'_1}(u_1)
\Phi_{\ve_1}(u_1)\ldots\Phi_{\ve_M}(u_M)
$$
in $\H_i$. The meaning of the arguments $u_j$ is the following. We
attach the same fixed value of the spectral parameter to each vertical
line and the zero spectral parameter to each horizontal line except
these $M$ lines where the values $-u_j$ live. For the homogeneous
eight-vertex model and the XYZ chain we have $u_1=\ldots=u_M=0$.

With these definitions the form factors are given by
$$
{}^{(i)}\langle0|\O|a_1\theta_1,\ldots,a_N\theta_N\rangle^{(i)}
={1\over\chi}
\Tr_{\H_i}\bigl(
\Psi^*_{a_N}(v_N)\ldots\Psi^*_{a_1}(v_1)\hat\O\rho
\bigr),
\qquad
\chi=\chi(x^4).
\eqlabel\EQformfactor
$$

Now consider the SOS model. Consider semi-infinite paths, {\it i.~e.}\
the sequences of heights $n_0,n_1,n_2,\ldots$, such that
$|n_k-n_{k+1}|=1$. Let $\H_{mn}$ be the space of formal linear
combinations of all semi-infinite paths such that $n_0=n$ and either
$n_{2k}=m$, $n_{2k+1}=m+1$ or $n_{2k}=m+1$, $n_{2k+1}=m$ depending on
the parity of $m-n$ for large enough $k$. Introduce the operators
$$
\rho_{mn}=[n]x^{4H_{mn}}:\H_{mn}\to\H_{mn},
\qquad
\Phi(u)^{mn'}_{mn}:\H_{mn}\to\H_{mn'},
\qquad
\Psi^*(u)^{m'}_m{}^n_n:\H_{mn}\to\H_{m'n}.
\eqlabel\EQsosctmvos
$$
Here $m'=m\pm1$, $n'=n\pm1$, {\it i.~e.}\ the pairs $(m',m)$ and $(n',n)$
are admissible. Below we always write simply $\Phi(u)^{n'}_n$ and
$\Psi^*(u)^{m'}_m$. The commutation relations read
$$
\eqalign{
\span
\Phi(u_1)^{n'}_s\Phi(u_2)^s_n
=\sum_{s'}\W(s,n;s',n'|u_1-u_2)
\Phi(u_2)^{n'}_{s'}\Phi(u_1)^{s'}_n,
\cr
\span
\sum_{s'}\tW(s',m;s,m'|u_1-u_2)
\Psi^*(u_1)^{m'}_{s'}\Psi^*(u_2)^{s'}_m
=\Psi^*(u_2)^{m'}_s\Psi^*(u_1)^s_m,
\cr
\span
\Phi(u_1)^{n'}_n\Psi^*(u_2)^{m'}_m
=\tau(u_1-u_2)\Psi^*(u_2)^{m'}_m\Phi(u_1)^{n'}_n
}\eqlabel\EQsoscommutation
$$
and
$$
\Phi(u)^{n'}_n{\rho_{mn}\over [n]}
={\rho_{mn'}\over [n']}\Phi(u-2)^{n'}_n,
\qquad
\Psi^*(u)^{m'}_m\rho_{mn}=\rho_{m'n}\Psi^*(u-2)^{m'}_m.
\eqlabel\EQsosdiff
$$
Here
$$
\tW(m_1,m_2;m_3,m_4|u)=-\W(m_1,m_2;m_3,m_4|u;\epsilon,r-1)
$$
The normalization conditions look like
$$
\eqalign{
\span
\sum_{n'}\Phi^*(u)^n_{n'}\Phi(u)^{n'}_n=1,
\qquad
\Phi(u)^{n''}_n\Phi^*(u)^n_{n'}=\delta_{n''n'},
\cr
\span
\Psi(u')^{m''}_m\Psi^*(u)^m_{m'}={1\over\pi}{\delta_{m''m'}\over u'-u}
+O((u'-u)^0)
}\eqlabel\EQsosnorm
$$
(admissibility of the pairs $(n,n')$, $(n,n'')$, $(m,m')$, and $(m,m'')$
is assumed) with
$$
\Phi^*(u)^{n'}_n=(-)^{n-m_0+1}(n'-n)[n]\Phi(u-1)^{n'}_n,
\qquad
\Psi(u)^{m'}_m=(-)^{m-m_0+1}{m'-m\overtext[m]'}\Psi^*(u-1)^{m'}_m.
\eqlabel\EQsosconj
$$
Here
$$
[u]'_i=h_i(u;r-1).
\eqlabel\EQbracketprime
$$

We want to incorporate the vertex-face correspondence into the
vertex operator algebra. For the type~I vertex operators it was done
in~\Refs{\LPone}. The result can be formulated as follows. There exist
two operators
$$
T(u_0)_{mn}:\H_{mn}\to\H_i,
\qquad
T(u_0)^{mn}:\H_i\to\H_{mn},
\qquad
i=n-m\pmod2,
\eqlabel\EQToperators
$$
such that
$$
[m]'\rho^{(i)}=\sum_{n\in2\Z+m+i}T(u_0)_{mn}\rho_{mn}T(u_0)^{mn}
\eqlabel\EQvfctm
$$
and
$$
\eqadvance\EQvfvo
\eqalign{
\sum_\ve t^*_\ve(u-u_0)_{n'}^n\Phi_\ve(u)T(u_0)_{mn}
&=T(u_0)_{mn'}\Phi(u)_n^{n'},
\cr
\sum_n t_\ve(u-u_0)^{n'}_n\Phi(u)^{n'}_nT(u_0)^{mn}
&=T(u_0)^{mn'}\Phi_\ve(u)
}\eqlabelno(\EQvfvo a)
$$
or, equivalently,
$$
\eqalign{
\Phi_\ve(u)T(u_0)_{mn}
&=\sum_{n'}t'_\ve(u-u_0)^{n'}_nT(u_0)_{mn'}\Phi(u)^{n'}_n,
\cr
\Phi(u)_n^{n'}T(u_0)^{mn}
&=\sum_{\ve}t^*_\ve(u-u_0)^n_{n'}T(u_0)^{mn'}\Phi_\ve(u).
}\eqlabelno(\EQvfvo a')
$$
It is straightforward to prove that (\EQvfctm), (\EQvfvo a) are consistent
with the algebras of $\Phi$ and $\rho$ operators for the eight-vertex and
SOS models (see Appendix~A).

As our aim is to express the objects of the eight-vertex model in terms of
the objects in the SOS model, which are considered to be well defined
due to the Lukyanov--Pugai free field representation, let us look at what
these relations mean for the SOS model. Consider the operator
$$
\Lambda(u_0)^{m'}_m{}^{n'}_n
=T(u_0)^{m'n'}T(u_0)_{mn}:\H_{mn}\to\H_{mn}.
\eqlabel\EQLambda
$$
In the case of vacuum expectation values, which correspond to
$N=0$ in (\EQformfactor), one can pull {\it e.~g.} $T(u_0)^{mn}$ to the
right by use of the cyclic property of trace and express
$\langle\O\rangle$ in terms of traces over $\H_{mn}$ of operators
$\Phi(u_j)^{n'}_n$, $\rho_{mn}$, and $\Lambda(u_0)^{mn'}_{mn}$:
$$
\eqalignno{
\langle\O\rangle^{(i)}
&=\sum_{n\in\Z+m+i}\sum_{\{n_k\},\{\nu_k\},n'}
t^*_{\ve'_M}(u_M-u_0)^n_{\nu_{M-1}}\ldots t^*_{\ve'_1}(u_1-u_0)^{\nu_1}_{n_0}
t_{\ve_1}(u_1-u_0)^{n_0}_{n_1}\ldots t_{\ve_M}(u_M-u_0)^{n_{M-1}}_{n'}
\cr
&\quad\times
{1\over\chi}\Tr_{\H_{mn}}(
\Phi^*(u_M)^n_{\nu_{M-1}}\ldots\Phi^*(u_1)^{\nu_1}_{n_0}
\Phi(u_1)^{n_0}_{n_1}\ldots\Phi(u_M)^{n_{M-1}}_{n'}
\Lambda(u_0)^m_m{}^{n'}_n\>{\rho_{mn}\over[m]'}).
\lnlabel\EQcorrfun}
$$
The operator $\Lambda$ is an infinite tail of the intertwining vectors that
cannot be cancelled if something nonintegrable is inserted into a finite
part of the lattice (Fig.~5).

\topinsert
%
%
\line{\hfil
\beginpicture
\setcoordinatesystem units <1cm,1cm> point at 0 0
\trl 0.0 1.0 , 3.0 1.0
\trl 0.0 0.0 , 3.0 0.0
\trl 4.0 1.0 , 6.0 1.0
\trl 4.0 0.0 , 6.0 0.0
\setdashes <3pt>
\trl 3.0 1.0 , 4.0 1.0
\trl 3.0 0.0 , 4.0 0.0
\trl 6.0 1.0 , 6.5 1.0
\trl 6.0 0.0 , 6.5 0.0
\setdashes <2.5pt>
\rl -0.5 0.5 , 6.5 0.5
\rl 0.5 1.5 , 0.5 1.0
\rl 0.5 0.0 , 0.5 -0.5
\setsolid
\ar -0.48 0.5 , -0.5 0.5
\ar 0.5 -0.48 , 0.5 -0.5
\ln 0.4 1.0 , 0.5 0.9
\ln 0.5 0.9 , 0.6 1.0
\ln 0.4 0.0 , 0.5 0.1
\ln 0.5 0.1 , 0.6 0.0
\rl 0.5 0.9 , 0.5 0.1
\ln 1.4 1.0 , 1.5 0.9
\ln 1.5 0.9 , 1.6 1.0
\ln 1.4 0.0 , 1.5 0.1
\ln 1.5 0.1 , 1.6 0.0
\rl 1.5 0.9 , 1.5 0.1
\ln 2.4 1.0 , 2.5 0.9
\ln 2.5 0.9 , 2.6 1.0
\ln 2.4 0.0 , 2.5 0.1
\ln 2.5 0.1 , 2.6 0.0
\rl 2.5 0.9 , 2.5 0.1
\ln 4.4 1.0 , 4.5 0.9
\ln 4.5 0.9 , 4.6 1.0
\ln 4.4 0.0 , 4.5 0.1
\ln 4.5 0.1 , 4.6 0.0
\rl 4.5 0.9 , 4.5 0.1
\ln 5.4 1.0 , 5.5 0.9
\ln 5.5 0.9 , 5.6 1.0
\ln 5.4 0.0 , 5.5 0.1
\ln 5.5 0.1 , 5.6 0.0
\rl 5.5 0.9 , 5.5 0.1
\put {$v_0$} [Br] at -0.6 0.45
\put {$u$} [tc] at 0.5 -0.6
\put {$n'$} [Bc] at 0.0 1.15
\put {$n$} [tc] at 0.0 -0.15
\put {$n'_1$} [Bc] at 1.0 1.15
\put {$n_1$} [tc] at 1.0 -0.15
\put {$n'_2$} [Bc] at 2.0 1.15
\put {$n_2$} [tc] at 2.0 -0.15
\put {$n'_3$} [Bc] at 3.0 1.15
\put {$n_3$} [tc] at 3.0 -0.15
\put {$m'$} [Bc] at 4.0 1.15
\put {$m$} [tc] at 4.0 -0.15
\put {$m'+1$} [Bc] at 5.0 1.15
\put {$m\smash{{}+1}$} [tc] at 5.0 -0.15
\put {$m'$} [Bc] at 6.0 1.15
\put {$m$} [tc] at 6.0 -0.15
\endpicture
\hfil}
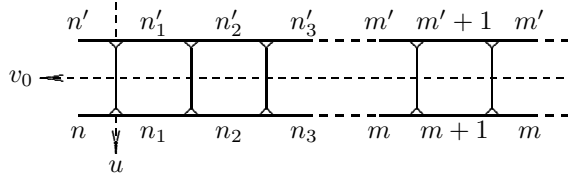
\figcap{5} The operator $\Lambda(u_0)^{m'}_m{}^{n'}_n$ ($u_0=u-v_0$).
The right solid part represents the condition at the infinity that
corresponds to the ground state sector $m$. The upper bold line forms
the operator $T(u_0)^{m'n'}$. The lower one forms $T(u_0)_{mn}$.\par
\endinsert

From (\EQvfvo a) we easily obtain
$$
\eqadvance\EQLambdavocommut
\Lambda(u_0)^{m'}_m{}^{n'}_s\Phi(u)^s_n
=\sum_{s'}\L(s,n;s',n'|u_0-u)\Phi(u)_{s'}^{n'}\Lambda(u_0)^{m'}_m{}^{s'}_n
\eqlabelno(\EQLambdavocommut a)
$$
with
$$
\L(n_1,n_2;n_3,n_4|u)
=\sum_\ve t^*_\ve(-u)^{n_2}_{n_1} t_\ve(-u)^{n_4}_{n_3}.
\eqlabel\EQLdef
$$
Explicitly,
$$
\eqalign{
\L(n,n\pm1;n'\pm1,n'|u)
&={[u\mp\half(n-n')][\half(n+n')]\over [u][n]},
\cr
\L(n,n\pm1;n'\mp1,n'|u)
&={[u\mp\half(n+n')][\half(n-n')]\over [u][n]},
}\eqlabel\EQLexpr
$$
for $n'-n\in2\Z$. Evidently,
$$
\L(n,n';n'',n|u)=\delta_{n'n''}.
\eqlabel\EQLdiag
$$

An important property of $\Lambda(u_0)$ is
$$
\Lambda(u_0)^{m'}_m{}^n_n=\delta_{m'm}.
\eqlabel\EQLambdadiag
$$
It follows directly from Fig.~5 and Fig.~3b. It means that the $\Lambda$
operator does not appear in the case $\O=1$. Moreover,
$$
\Lambda(u_0)^{m'}_m{}^{n'}_n=0
\hbox{\quad if\quad $m'<m$, $n'>n$ or $m'>m$, $n'<n$}.
\eqlabel\EQLambdagreaterless
$$
Indeed, if $\Lambda$ is nonzero for, {\it e.~g.}, $n'>n$ and $m'<m$,
there must exist a point $j$ on the half-infinite chain where
$n'_j=n_j$ and therefore $n'_k=n_k$ for all $k>j$. Hence, $m'=m$ which
contradicts to the initial assumption.

Our next task is to extend the algebra (\EQvfctm), (\EQvfvo a) to the $\Psi$
operators. Let $\tdt_\ve(u)^{m'}_m$ be, up to a scalar factor,
$t_\ve(u)^{m'}_m$ after substitution $r\to r'=r-1$. More precisely,
$$
\eqadvance\EQintertwiningtilde
\tdt_+(u)^{m'}_m=\tf(u)
\theta_3\!\left({(m'-m)u+m'\over2r'};\i{\pi\over2\epsilon r'}\right),
\qquad
\tdt_-(u)^{m'}_m=(-1)^{m-m_0+1}\tf(u)
\theta_4\!\left({(m'-m)u+m'\over2r'};\i{\pi\over2\epsilon r'}\right)
\eqlabelno(\EQintertwiningtilde a)
$$
with $\tf(u)$ satisfying the equation
$$
[u]'\tf(u)\tf(u-1)=C'\equiv{[0]^{\prime2}_4\over
2\theta_3(0;\i\pi/2\epsilon r')\theta_4(0;\i\pi/2\epsilon r')}.
\eqlabel\EQCprimedef
$$
This equation has an infinite number of solutions.
The relevant solution of this equation will be chosen in Sec.~5
after discussing the free field representation and an explicit calculation
of a form factor in a particular case. Here we cite the answer:
$$
\eqalign{
\tf(u)
&=\i\sqrt{C'}x^{-u^2/2(r-1)-ru/2(r-1)-1/4}\tf_1(x^{2u}),
\cr
\tf_1(z)
&={1\over\sqrt{(x^{2r-2};x^{2r-2})_\infty}}
{(x^{2r+2}z;x^4,x^{2r-2})_\infty(x^2z^{-1};x^4,x^{2r-2})_\infty
\over(x^{2r}z;x^4,x^{2r-2})_\infty(z^{-1};x^4,x^{2r-2})_\infty}.
}\eqlabelno(\EQintertwiningtilde b)
$$

Consider the algebra
$$
\eqalign{
\Psi^*_\ve(u)T_{mn}(u_0)
&=\sum_{m'} T_{m'n}(u_0)\Psi^*(u)^{m'}_m
\tdt^*_\ve(u-u_0-\Delta u_0)^m_{m'},
\cr
\Psi^*(u)^{m'}_mT^{mn}(u_0)
&=\sum_\ve T^{m'n}(u_0)\Psi^*_\ve(u)\tdt_\ve(u-u_0-\Delta u_0)^{m'}_m
}\eqlabelno(\EQvfvo b)
$$
or, equivalently,
$$
\eqalign{
\sum_\ve\Psi^*_\ve(u)T_{mn}(u_0)\tdt_\ve(u-u_0-\Delta u_0)^{m'}_m
&=T_{m'n}(u_0)\Psi^*(u)^{m'}_m,
\cr
\sum_m\Psi^*(u)^{m'}_mT^{mn}(u_0)\tdt^*_\ve(u-u_0-\Delta u_0)^m_{m'}
&=T^{m'n}(u_0)\Psi^*_\ve(u).
}\eqlabelno(\EQvfvo b')
$$
This algebra is consistent with (\EQevcommutation--\skipsec\EQevnorm),
(\EQsoscommutation--\skipsec\EQsosnorm), (\EQvfvo a) for any value of
$\Delta u_0$ subject to (\EQvfctm) (see Appendix~A). The
value of $\Delta u_0$ will be fixed by the free field representation in
Sec.~4. Now rewrite these equations in terms of $\Lambda(u_0)$:
$$
\Psi^*(u)_s^{m'}\Lambda(u_0)^s_m{}^{n'}_n
=\sum_{s'}\Lambda(u_0)^{m'}_{s'}{}^{n'}_n\Psi^*(u)_m^{s'}
\tL(s',m;s,m'|u_0+\Delta u_0-u),
\eqlabelno(\EQLambdavocommut b)
$$
where
$$
\tL(m_1,m_2;m_3,m_4|u)
=\sum_\ve\tdt^*_\ve(-u)^{m_2}_{m_1}\tdt_\ve(-u)^{m_4}_{m_3}
=\L(m_1,m_2;m_3,m_4|u;\epsilon,r-1).
\eqlabel\EQtLdef
$$

To stress the difference between commutation relations
(\EQLambdavocommut a) and (\EQLambdavocommut b) let us write down the
simplest of them explicitly. For $n'=s$ in (\EQLambdavocommut a) using
(\EQLdiag) we obtain the identity
$$
\Lambda(u_0)^{m'}_m{}^{n'}_{n'}\Phi(u)^{n'}_n
=\Phi(u)^{n'}_n\Lambda(u_0)^{m'}_m{}^n_n,
$$
which is trivial due to (\EQLambdadiag). Now let $n'<n$. For $s=m$ in
(\EQLambdavocommut b) we have
$$
\eqadvance\EQPsiLsimple
\eqalignno{
\Psi^*(u)^{m+1}_m\Lambda(u_0)^m_m{}^{n'}_n
&=\Lambda(u_0)^{m+1}_{m+1}{}^{n'}_n\Psi^*(u)^{m+1}_m,
\lnlabelno(\EQPsiLsimple a)
\cr
\Psi^*(u)^{m-1}_m\Lambda(u_0)^m_m{}^{n'}_n
&=\Lambda(u_0)^{m-1}_{m-1}{}^{n'}_n\Psi^*(u)^{m-1}_m
\cr
&\quad
+\Lambda(u_0)^{m-1}_{m+1}{}^{n'}_n\Psi^*(u)^{m+1}_m
{[u_0+\Delta u_0-u+m]'[1]'\overtext[u_0+\Delta u_0-u]'[m+1]'}.
\lnlabelno(\EQPsiLsimple b)}
$$
These simplest commutation relations turn out to be nontrivial. They
mean that the diagonal in $m$ elements of $\Lambda$ commute with one
component of $\Psi^*(u)$, while the commutator with the second one
contains an off-diagonal in $m$ matrix element of $\Lambda$. Below, the
identity (\EQPsiLsimple b) will give us a hint how to find the free field
realization of the off-diagonal elements of the $\Lambda$ operator.

\nsec Free field representation

Recall the free field representation of the SOS model\refs{\LukPug}.
Consider a Heisenberg algebra of operators $a_k$ with nonzero integer
$k$ and a pair of `zero-mode' operators $\P$ and $\Q$ with the
commutation relations
$$
[\P,\Q]=-\i,\qquad
[a_k,a_l]=k{\[k\]_x\[(r-1)k\]_x\over\[2k\]_x\[rk\]_x}\delta_{k+l,0}
\quad\hbox{with}\quad
\[u\]_x={x^u-x^{-u}\over x-x^{-1}}.
\eqlabel\EQheisenberg
$$
The `$q$-number' $\[u\]_x$ here should not be confused with the `elliptic
$q$-numbers' $[u]_i$ ($i=1,\ldots,4$). It is also useful to introduce
the operators
$$
\tilde a_k={\[rk\]_x\over\[(r-1)k\]_x}a_k.
\eqlabel\EQatilde
$$
The normal ordering operation $\lcolon\ldots\rcolon$
places $\P$ to the right of $\Q$ and $a_k$ with positive $k$
to the right of $a_{-k}$. It will be convenient to assign
$$
\alpha_+=\sqrt{a_+}=\sqrt{r\over r-1},
\qquad
\alpha_-=-\sqrt{a_-}=-\sqrt{r-1\over r},
\qquad
2\alpha_0=\alpha_++\alpha_-={1\over\sqrt{r(r-1)}}.
\eqlabel\EQalphapmdef
$$
Now introduce the fields
$$
\eqalign{
\varphi(z)
&={\alpha_-\over\sqrt2}(\Q-\i\P\log z)-\sum_{k\neq0}{a_k\over\i k}z^{-k}.
\cr
\tilde\varphi(z)
&={\alpha_+\over\sqrt2}(\Q-\i\P\log z)+\sum_{k\neq0}{\tilde a_k\over\i k}z^{-k}.
}\eqlabel\EQvarphidef
$$
These fields enter the exponential operators
$$
\eqalign{
V(u)
&=z^{(r-1)/4r}\lcolon\e^{\i\varphi(z)}\rcolon,
&\qquad
\bV(u)
&=z^{(r-1)/r}\lcolon\e^{-\i\varphi(x^{-1}z)-\i\varphi(xz)}\rcolon,
\cr
\tV(u)
&=z^{r/4(r-1)}\lcolon\e^{\i\tilde\varphi(z)}\rcolon,
&\qquad
\tbV(u)
&=z^{r/(r-1)}\lcolon\e^{-\i\tilde\varphi(x^{-1}z)-\i\tilde\varphi(xz)}\rcolon,
}\eqlabel\EQVbarVdef
$$
and Lukyanov's screening operators
$$
\eqalign{
x(u,C)
&={\epsilon\over\eta}\int_C{dv\over\i\pi}\,
\bV(v){[v-u+\half-\sqrt{2r(r-1)}\,\P]
\over[v-u-\half]},
\cr
\tilde x(u,C)
&={\epsilon\over\eta'}\int_C{dv\over\i\pi}\,
\tbV(v){[v-u-\half+\sqrt{\textstyle2r(r-1)}\,\P]'
\overtext[v-u+\half]'}.
}\eqlabel\EQscreening
$$
\eqadvance\EQetadef
\eqadvance\EQXYscreening
\eqadvance\EQYXcorresp
\eqadvance\EQPvacdef
\eqadvance\EQbosvertex
The constants $\eta$, $\eta'$ are defined by the normalization conditions
(\EQsosnorm) together with the definitions (\EQbosvertex) below and read
$$
\eqalign{
\eta^{-1}
&=\i[1]\,x^{r-1\over2r}
{(x^2;x^{2r})_\infty\over(x^{2r-2};x^{2r})_\infty}
{(x^6;x^4,x^{2r})_\infty(x^{2r+2};x^4,x^{2r})_\infty
\over
(x^4;x^4,x^{2r})_\infty(x^{2r+4};x^4,x^{2r})_\infty},
\cr
\eta^{\prime-1}
&=-{2\epsilon\over\pi}[1]'x^{-{r\over2(r-1)}}
{(x^{2r-2};x^{2r-2})^2_\infty\over(x^{2r};x^{2r-2})^2_\infty}
{(x^4;x^4,x^{2r-2})_\infty(x^{2r+2};x^4,x^{2r-2})_\infty
\over
(x^2;x^4,x^{2r-2})_\infty(x^{2r+4};x^4,x^{2r-2})_\infty}.
}\eqlabelno(\EQetadef)
$$
Now let us fix the contours. Let $C^-_u$ and $C^+_u$ go from
$u-{\i\pi\over2\epsilon}$ to $u+{\i\pi\over2\epsilon}$ to the left
and to the right of $u$ respectively.%
\nfootnote{We assume that the contours $C^\pm_u$ go to the left of all
poles in the `main rectangle' related to the operators that are to the
right of the screening operator and to the right of all poles related
to the operators placed to the left of the screening operators. The
`main rectangle' is understood as a rectangle with sides $r$ along the
real axis and $\pi\over\epsilon$ along the imaginary axis that contains
all points $u_i$, $v_i$ {\it etc.} It is well defined for large enough
$r$ and for points $u_i$, $v_i$,\dots\ close enough to each other. In
the gereral case the operator products are considered as analytic
continuation from this region.}
Then
$$
\eqalign{
X(u)
&=x(u,C^-_{u+1/2}),
&\qquad
Y(u)
&=x(u-1,C^+_{u-1/2}),
\cr
\tX(u)
&=\tilde x(u,C^-_{u-1/2}),
&\qquad
\tY(u)
&=\tilde x(u+1,C^+_{u+1/2}).
}\eqlabelno(\EQXYscreening)
$$
These operators satisfy the equations
$$
Y(u)V(u)=V(u)X(u),
\qquad
\tY(u)\tV(u)=\tV(u)\tX(u).
\eqlabelno(\EQYXcorresp)
$$

Define the Fock spaces $\F_{mn}$ generated by the operators $a_{-k}$
($k>0$) from the highest weight vectors $|P_{mn}\rangle$ such that
$$
a_k|P_{mn}\rangle=0\quad (k>0),
\qquad
\P|P_{mn}\rangle=P_{mn}|P_{mn}\rangle,
\qquad
P_{mn}={1\over\sqrt2}(\alpha_+m+\alpha_-n).
\eqlabelno(\EQPvacdef)
$$
There are strong evidences that $\F_{mn}$ can be identified with $\H_{mn}$
for generic $r$.

The vertex operators are defined on $\F_{mn}$ as follows:
$$
\eqalign{
\Phi(u)^{n+1}_n
&={\i^{m-n}\over[n]}V(u),
\cr
\Phi(u)^{n-1}_n
&=(-)^{n-m_0+1}{\i^{m-n}\over[n]}V(u)X(u),
\cr
\Psi^*(u)^{m+1}_m
&=\tV(u),
\cr
\Psi^*(u)^{m-1}_m
&=(-)^{n-m_0}\tY(u)\tV(u).
}\eqlabelno(\EQbosvertex)
$$
The corner Hamiltonian $H_{mn}$ is the restriction to $\F_{mn}$ of the
operator
$$
H={\P^2\over2}+\sum_{k=1}^\infty
{\[2k\]_x\[rk\]_x\over\[k\]_x\[(r-1)k\]_x}a_{-k}a_k.
\eqlabel\EQbosH
$$

Later we shall need the following fact on the SOS model. The weights
(\EQWmatrix) are invariant under the substitution $n_k\to -n_k$. So the
model is invariant under the change
$$
n_k\to-n_k,
\qquad
m\to-m,
\qquad
m_0\to-m_0.
\eqlabel\EQchangesigns
$$
The free field representation (\EQbosvertex) is not invariant under
this substitution. So there exists another free field representation,
where $\H_{mn}$ is identified with $\F_{-m\>-n}$:
$$
\eqalign{
\Phi(u)^{n+1}_n
&=(-)^{m-m_0}{\i^{m-n}\over[n]}V(u)X(u),
\cr
\Phi(u)^{n-1}_n
&=(-)^{m-n+1}{\i^{m-n}\over[n]}V(u),
\cr
\Psi^*(u)^{m+1}_m
&=(-)^{n-m_0}\tY(u)\tV(u),
\cr
\Psi^*(u)^{m-1}_m
&=\tV(u).
}\eqlabel\EQbosvertexsecond
$$
Of course, this representation gives the same multipoint local height
probabilities as the first representation.

Our task is to realize the operator $\Lambda(u)$ in terms of free
fields. The elements diagonal in $m$ are known\refs{\LPone}. For
$n'=n-2l\le n$ they read
$$
\eqadvance\EQbosLambdadiag
\left.\Lambda(u)^m_m{\,}^{n-2l}_n\right|_{\F_{mn}}
=(-)^{(n-m_0+1)l}{[n-2l]\over [n]}X^l(u)
\quad\hbox{for $l\ge0$.}
\eqlabelno(\EQbosLambdadiag a)
$$
The elements with $n'>n$ have no realization in this representation.
However, the operator $\Lambda$ enters each trace once. Therefore,
one can apply the second representation (\EQbosvertexsecond) to
the traces that contain $\Lambda$ operators with $n'>n$
so that
$$
\left.\Lambda(u)^m_m{\,}^{n+2l}_n\right|_{\F_{-m\,-n}}
=(-)^{(n-m_0+1)l}{[n+2l]\over [n]}X^l(u)
\quad\hbox{for $l\ge0$.}
\eqlabelno(\EQbosLambdadiag b)
$$

Let us return to the equations (\EQPsiLsimple). In the free field
representation, the equation (\EQPsiLsimple a) is an evident
consequence of commutativity of $\tV(u)$ and $X(u_0)$. The equation
(\EQPsiLsimple b) is less trivial. For $n'=n-2$ it can be written as
follows
$$
\Lambda(u_0)^{m-2}_m{\,}^{n-2}_n\tV(u)
=-{[u_0+\Delta u_0-u]'[m]'\overtext[u_0+\Delta u_0-u+m-1]'[1]'}
{[n-2]\over[n]}
\,[\tY(u),X(u_0)]\,\tV(u).
$$
It can be expected that the operator $\Lambda(u_0)^{m-2}_m{\,}^{n-2}_n$
is more or less proportional to the commutator of $\tY$ and $X$.
This commutator is calculated in Appendix~C. The answer is
$$
\eqadvance\EQYXcommutator
\eqalignno{
[\tY(u'),X(u)]|_{\F_{mn}}
&={\epsilon\over\eta\eta'\sh\epsilon}\biggl(
{[m-1]'\overtext\partial[0]'}
{[u'-u+\half-n]\over[u'-u-\half]}
\,W_+(u')
\cr
&\quad
+{[n-1]\over\partial[0]}
{[u'-u+{3\over2}-m]'\overtext[u'-u+\half]'}
\,W_-(u)\biggr).
\lnlabelno(\EQYXcommutator a)}
$$
Here $\partial[0]=d[u]/du|_{u=0}$,
$\partial[0]'=d[u]'/du|_{u=0}$ and
$$
W_+(u)=W(u+{\textstyle{r\over2}}),
\qquad
W_-(u)=W(u-{\textstyle{r-1\over2}})
\eqlabelno(\EQYXcommutator b)
$$
with
$$
\eqalign{
W(u)
&=z^{1/r(r-1)}\lcolon\e^{\i\varphi_0(z)}\rcolon,
\cr
\varphi_0(z)
&=-2\sqrt2\alpha_0(\Q-\i\P\log z)
-\sum_{k\ne0}{\[2k\]_x\over\[(r-1)k\]_x}{a_k\over\i k}z^{-k}.
}\eqlabel\EQWoperator
$$
It follows from the identities of Appendix~B that
$$
W_+(u)\tV(u)=0.
$$
It means that the first term in (\EQYXcommutator) does not contribute
to $\Lambda(u)^{m-2}_m{\,}^{n-2}_n$, and the expression
$$
\Lambda(u)^{m-2}_m{\,}^{n-2}_n
=-{\epsilon\over\eta\eta'\sh\epsilon}
{[m]'\overtext[1]'}{[n-1][n-2]\over\partial[0]\,[n]}
\,W_-(u)
\eqlabel\EQLambdasimplest
$$
is consistent with (\EQPsiLsimple) for
$$
\Delta u_0=-1/2.
\eqlabel\EQDeltauzero
$$
Consider the limit $u'\to u-\half$ in (\EQYXcommutator a). It is singular.
It can be seen that the product $X(u)\tY(u')$ is regular in this limit,
while the product $\tY(u')X(u)$ has a pole because of pinching contours
between poles. It means that
$$
\Lambda(u)^{m-2}_m{\,}^{n-2}_n
\sim\lim_{u'\to u}[u'-u]'\tY(u'-{\textstyle\half})X(u).
\eqlabel\EQYXprodlim
$$
Generally, we conjecture that
$$
\Lambda(u_0)^{m-2k}_m{\,}^{n-2l}_n
\sim
\lim_{u'\to u}[u'-u]'\tY^k(u'-{\textstyle\half})X^l(u).
\eqlabel\EQLambdaconj
$$
It is shown in Appendix~C that the r.~h.~s.\ is well-defined and
given by
$$
\eqalignno{
&\lim_{u'\to u}[u'-u]'\tY^k(u'-{\textstyle\half})X^l(u)
\cr
&\qquad\qquad
=(-)^{k+l+1}{\epsilon\over\eta\eta'\sh\epsilon}
{[k]'[m-k]'\overtext[1]'}
{[l][n-l]\over [1]\,\partial[0]}
\tX^{k-1}(u-{\textstyle{1\over2}})
W_-(u)
Y^{l-1}(u).
\lnlabel\EQlimformula}
$$
It is important that the contours in $\tX^{k-1}$ go to the left
of all poles while the contours in $Y^{l-1}$ go to the right of all
poles.

So we have
$$
\eqadvance\EQLambdafin
\eqalignno{
\left.\Lambda(u)^{m-2k}_m{\,}^{n-2l}_n\right|_{\F_{mn}}
&=C^{m-2k}_m{\,}^{n-2l}_n
\lim_{u'\to u}[u'-u]'\tY^k(u'-{\textstyle\half})X^l(u)
\cr
&=D^{m-2k}_m{\,}^{n-2l}_n\tX^{k-1}(u-{\textstyle{1\over2}})
W_-(u)
Y^{l-1}(u),
\lnlabelno(\EQLambdafin a)
\cr
\left.\Lambda(u)^{m+2k}_m{\,}^{n+2l}_n\right|_{\F_{-m\,-n}}
&=C^{m+2k}_m{\,}^{n+2l}_n
\lim_{u'\to u}[u'-u]'\tY^k(u'-{\textstyle\half})X^l(u)
\cr
&=D^{m+2k}_m{\,}^{n+2l}_n\tX^{k-1}(u-{\textstyle{1\over2}})
W_-(u)
Y^{l-1}(u),
\lnlabelno(\EQLambdafin b)}
$$
for $k,l>1$, with some coefficients
$C^{m'}_m{\,}^{n'}_n$ and $D^{m'}_m{\,}^{n'}_n$. To find the
coefficients let us turn to the commutation
relations~(\EQLambdavocommut b). Their explicit form for $\Delta
u_0=-1/2$ is written out in Appendix~D. It is easy to find from them
that
$$
\eqalignno{
&\Psi^*(u)^{m-2k+1}_{m-2k}\Lambda(u_0)^{m-2k}_m{\,}^{n-2l}_n
-{[u_0-u-m+k-\half]'[k-1]'\overtext[u_0-u-k+\half]'[m-k]'}
\Psi^*(u)^{m-2k+1}_{m-2k+2}\Lambda(u_0)^{m-2k+2}_m{\,}^{n-2l}_n
\cr
&\quad
={[u_0-u+\half]'[m]'[m-2k+1]'\overtext[u_0-u-k+\half]'[m+1]'[m-k]'}
\Lambda(u_0)^{m-2k+1}_{m+1}{\,}^{n-2l}_n\Psi^*(u)^{m+1}_m.
}
$$
Note that $\tV(u)W_(u_0)$ has a zero at $u=u_0+\half$ and a pole at
$u=u_0-\half$. Hence, both sides of the equation vanish at $u=u_0+\half$
and this point is not interesting. But at $u=u_0-\half$ both terms of
the l.~h.~s.\ have a pole while the r.~h.~s. remains finite. It means
that near this point we may write
$$
\Psi^*(u)^{m-2k+1}_{m-2k}\Lambda(u_0)^{m-2k}_m{\,}^{n-2l}_n
=\Psi^*(u)^{m-2k+1}_{m-2k+2}\Lambda(u_0)^{m-2k+2}_m{\,}^{n-2l}_n+O(1),
\qquad
u\to u_0-\half.
$$
By use of (\EQbosvertex) and (\EQYXcorresp) we obtain
$$
\tV(u)\Lambda(u_0)^{m-2k}_m{\,}^{n-2l}_n
=(-)^{n-m_0}\tV(u)\tX(u_0-\texthalf)\Lambda(u_0)^{m-2k+2}_m{\,}^{n-2l}_n
+O(1),
\qquad
u\to u_0-\half.
$$
Here we used that the product $\tX(u)\Lambda(u_0)$ is regular at this
point. Hence,
$$
\Lambda(u_0)^{m-2k}_m{\,}^{n-2l}_n
=(-)^{n-m_0}\tX(u_0-\texthalf)\Lambda(u_0)^{m-2k+2}_m{\,}^{n-2l}_n.
$$
Similarly, from (\EQLambdavocommut a) and the identity $V(u)W_-(u)=0$
we obtain
$$
\Lambda(u_0)^{m-2k}_m{\,}^{n-2l}_n
=(-)^{n-m_0}{[l][n][n-l]\over[l-1][n-2][n-l-1]}
\Lambda(u_0)^{m-2k}_m{\,}^{n-2l}_{n-2}Y(u_0).
$$
Together with (\EQLambdasimplest) it defines the
$D$ and, therefore, $C$ coeffitients
$$
\eqalign{
D^{m-2k}_m{\,}^{n-2l}_n
&=(-)^{(n-m_0)(k+l)+1}
{\epsilon\over\eta\eta'\sh\epsilon}
{[m]'\overtext[1]'}
{[l][n-l][n-2l]\over\partial[0]\,[1][n]}.
\cr
C^{m-2k}_m{\,}^{n-2l}_n
&=(-)^{(n-m_0+1)(k+l)}{[m]'\overtext[k]'[m-k]'}
{[n-2l]\over [n]}.
}\eqlabel\EQLambdacoeffs
$$
These expressions are valid for $k,l<0$ as well. Indeed, for $m'>m$,
$n'>n$ we have {\it e.~g.}\
$D^{m'}_m{}^{n'}_n=D^{-m'}_{-m}{}^{-n'}_{-n}|_{m_0\to-m_0}$ by
definition of the second representation. It is evident that $D$
coefficients defined by (\EQLambdacoeffs) satisfy this equation.

From (\EQLambdadiag) and (\EQLambdagreaterless) we know that
$\Lambda(u_0)^{m\pm2k}_m{\,}^n_n
=\Lambda(u_0)^{m\pm2k}_m{\,}^{n\mp2l}_n=0$ for $k,l>0$,
while the elements $\Lambda(u_0)^m_m{\,}^{n\pm2l}_n$ ($l\ge0$) are given
by~(\EQbosLambdadiag). It means that we never encounter the situation
when an element of the $\Lambda$ operator could not be realized
in either the first or the second bosonic representation.

The formulas (\EQLambdafin), (\EQLambdacoeffs) are the main result of
the paper. To prove them one has to check the commutation relation
(\EQLambdavocommut) with $\Lambda(u)$ from~(\EQLambdafin). It is
enough to check it for (\EQLambdavocommut a).
This is a more or less direct but rather cumbersome
calculation, and basically the same as the proof of (\EQLambdadiag) by
substituting it into~(\EQLambdavocommut a), given in the Appendix~C
of~\Refs{\LPone}, so we only give guidelines of it in Appendix~D.
Eq.~(\EQLambdavocommut b) with $m'<m$ follows directly from that with
$m'=m$, which was checked in~\Refs{\LPone}, and~(\EQLambdaconj).

The only thing that has not yet been
fixed is the function $\tf(z)$ that enter the intertwining vectors for
excitations. To fix it we need to evaluate, at least, the simplest
form factor.

\nsec Two-particle form factor of the $\sigma^z$ operator

Consider the operator
$\sigma^z$. Its free field realization is given by%
\nfootnote{Recall that $u=0$ for the homogeneous model.}
$$
\widehat{\sigma^z}=\sum_\ve \ve\Phi_{-\ve}(u-1)\Phi_\ve(u).
\eqlabel\EQsigmazhat
$$
It means that the two-particle form factor of this operator is given by
$$
F_{a_1a_2}^{(i)}(\theta_1,\theta_2)
\equiv{}^{(i)}\langle0|\sigma^z|a_1\theta_1,a_2\theta_2\rangle
={1\over\chi}
\sum_\ve \ve\Tr_{\H_i}(\Psi^*_{a_2}(v_2)\Psi^*_{a_1}(v_1)
\Phi_{-\ve}(u-1)\Phi_\ve(u)\rho)
\eqlabel\EQsigmazffdef
$$
with $\theta_j=\pi^2/2\epsilon+\i\pi v_j$. The direct way to calculate
it is to substitute (\EQvfctm) with arbitrary $m$ into the r.~h.~s.\ of
(\EQsigmazffdef) and using (\EQvfvo) to reduce the trace to a linear
combination of the traces for the SOS model:
$$
F_{a_1a_2}^{(i)}(\theta_1,\theta_2)
=\sum_{m_1,m_2}
F^{(i)}_{m_1m_2m}(\theta_1,\theta_2)
\tdt^*_{a_2}(v_2-u_0+\texthalf)^m_{m_1}
\tdt^*_{a_1}(v_1-u_0+\texthalf)^{m_1}_{m_2}
$$
with
$$
\eqalignno{
&F_{m_1m_2m}^{(i)}(\theta_1,\theta_2)
\cr
&\quad
={1\over\chi}\sum_\ve\sum_{n\in2\Z+m+i}
\ve\Tr_{\H_{mn}}
\Bigl(\Psi^*(v_2)^m_{m_2}\Psi^*(v_1)^{m_2}_{m_1}
T(u_0)^{m_2n}\Phi_{-\ve}(u-1)\Phi_\ve(u)T(u_0)_{mn}
{\rho_{mn}\overtext[m]'}\Bigr)
\lnlabel\EQsigmazsosffdef
\cr
&\quad
={1\over\chi}\sum_\ve\sum_{n\in2\Z+m+i}\sum_{n_1n_2}
\ve t_{-\ve}(u-u_0-1)^n_{n_2}t_\ve(u-u_0)^{n_2}_{n_1}
\cr
&\qquad\times
\Tr_{\H_{mn}}
\Bigl(\Psi^*(v_2)^m_{m_2}\Psi^*(v_1)^{m_2}_{m_1}
\Phi(u-1)^n_{n_2}\Phi(u)^{n_2}_{n_1}
\Lambda(u_0)^{m_1}_m{\,}^{n_1}_n{\rho_{mn}\overtext[m]'}\Bigr).
\lnlabel\EQsigmazffcalc}
$$
Now we can apply the bosonization technique, but we prefer to use the
following trick. By pulling $T(u_0)^{m_1n}$ in (\EQsigmazsosffdef) to the
left by use of (\EQvfvo b) we obtain
$$
F_{m_1m_2m}^{(i)}(\theta_1,\theta_2)
=\sum_{a_1a_2}F_{a_1a_2}^{(i)}(\theta_1,\theta_2)
\tdt_{a_2}(v_2-u_0+\texthalf)^m_{m_2}
\tdt_{a_1}(v_1-u_0+\texthalf)^{m_2}_{m_1}.
\eqlabel\EQsigmazffrel
$$
The idea is to fix $m_1-m$, $m_2-m$ and to consider (\EQsigmazffrel) as
a {\it functional equation\/} for the form factor
$F_{a_1a_2}^{(i)}(\theta_1,\theta_2)$ considered as an analytic
function of variables $\theta_1$, $\theta_2$, $m$. The requirement of
$m$-independence of $F_{a_1a_2}$ fixes the solution uniquely. Indeed,
consider the products
$\tdt_{a_1}(v_1)^m_{m_1}\tdt_{a_2}(v_2)^m_{m_2}/(a_1a_2)^{m-m_0}f(v_1)f(v_2)$
as theta-functions of the variable $w=m/2r'$ with the quasiperiods $1$ and
$\i\pi/\epsilon r'$. It is extracted from the periodicity properties
that they belong to a four-dimensional space. Moreover, as
$\tdt_+/\tdt_-$ is not a constant in $w$, these products form a basis in
this space for generic values of $v_1$, $v_2$.

Let us, for example, fix
$$
m_2+1=m_1+2=m.
$$
Then the only terms in (\EQsigmazffcalc) that make a nonzero contribution
are those with
$$
n_2+1=n_1+2=n.
$$
It means that
$$
\eqalignno{
F_{m-2\,m-1\,m}^{(i)}(\theta_1,\theta_2)
&={1\over\chi}\sum_\ve\sum_{n\in2\Z+m+i}
\ve t_{-\ve}(u-u_0-1)^n_{n-1}t_\ve(u-u_0)^{n-1}_{n-2}
\cr
&\quad\times
\Tr_{\H_{mn}}
\Bigl(\Psi^*(v_2)^m_{m-1}\Psi^*(v_1)^{m-1}_{m-2}
\Phi(u-1)^n_{n-1}\Phi(u)^{n-1}_{n-2}
\Lambda(u_0)^{m-2}_m{\,}^{n-2}_n{\rho_{mn}\overtext[m]'}\Bigr).
\lnlabel\EQsigmazffcalcspec}
$$
But $\Lambda(u_0)^{m-2}_m{\,}^{n-2}_n$ is expressed in terms of the
field $W_-(u_0)$ and contains no integrations. The $\Phi$ and $\Psi^*$
fields in this trace contain no integrations as well. It means that it
is possible to obtain an explicit representation for the two-point form
factor of $\sigma^z$ in terms of infinite products and elliptic
functions only, without any integrations.

Let us make some remarks on traces of the vertex operators. Consider
any operator $U(u)$ of the form
$$
U(u)=U^0(u)\e^{\phi^+(u)}\e^{\phi^-(u)},
\eqlabel\EQvosplit
$$
where $U^0(u)$ is the zero-mode operator of the form
$z^{\alpha^2/2}\e^{i\alpha\Q}z^{\alpha\P}$, and operators
$\phi^\pm(u)$ are positive/negative frequency parts of
the form $\sum_{k=1}^\infty A_{\pm k}a_{\mp k}z^{\pm k}$.

Denote by $\Tr_*$ the trace over oscillator modes and by $H^*$ the
oscillator contribution to $H_{mn}$. Besides, let
$$
\chi^*=\Tr_*(x^{4H^*})={1\over(x^4;x^4)_\infty}.
\eqlabel\EQchistardef
$$
Then
$$
\Tr_{\F_{mn}}(U_N(u_N)\ldots U_1(u_1))
=\langle P_{mn}|U^0_N(u_N)\ldots U^0_1(u_1)|P_{mn}\rangle\,
\chi^*\prod_{i=1}^N c_i\prod_{i<j}g_{ij}(u_i-u_j)
\eqlabel\EQtrsplit
$$
with
$$
\eqalign{
\log c_i
&={1\over\chi^*}\Tr_*(\phi^+_i(0)\phi^-_i(0)x^{4H^*}),
\cr
\log g_{ij}(u)
&={1\over\chi^*}\Tr_*(\phi_i(0)\phi_j(u)x^{4H^*}),
\qquad
\phi_i(u)=\phi^+_i(u)+\phi^-_i(u).
}\eqlabel\EQcgdef
$$
For the vertex operators used here the expressions for $c_i$ and
$g_{ij}$ are listed in the Appendix~E.

By substituting (\EQbosvertex), (\EQLambdafin a), and (\EQLambdacoeffs)
into (\EQsigmazffcalcspec) we obtain
$$
\eqalign{
F_{m-2\,m-1\,m}^{(i)}(\theta_1,\theta_2)
&=(-)^{m-m_0}{c\over\chi}\sum_{n\in2\Z+m+i}K_nT_{mn},
\cr
T_{mn}
&=\Tr_{\H_{mn}}(\tV(v_2)\tV(v_1)V(u-1)V(u)
W_-(u_0)x^{4H}),
}\eqlabel\EQsigmazffvvvv
$$
where
$c=\i\epsilon/([1]'\,\partial[0]\eta\eta'\sh\epsilon)$
and
$$
K_n
=(-)^{n-m_0}\sum_\ve \ve\,t_{-\ve}(u-u_0-1)^n_{n-1}t_\ve(u-u_0)^{n-1}_{n-2}
={[n-1+u-u_0]_4[0]_4\over [u-u_0]_1}.
\eqlabel\EQKppdef
$$
The trace $T_{mn}$ can be factorized as
$$
T_{mn}=h_{mn}(u,u_0;v_1,v_2)g(u,u_0;v_1,v_2).
$$
Here $h_{mn}$ is the zero-mode contribution:
$$
\eqalignno{
h_{mn}(u,u_0;v_1,v_2)
&=\langle P_{mn}|\tV^0(v_2)\tV^0(v_1)V^0(u-1)V^0(u)W^0_-(u_0)
x^{2\P^2}|P_{mn}\rangle
\cr
&=\e^{-\epsilon a_+m^2-2\epsilon(a_+-2-(a_+-1)u_0-u+{a_+\over2}(v_1+v_2))m
+\epsilon(2a_++a_--1-2(2-a_-)u+{a_+\over2}(3v_1+v_2))}
\cr
&\quad\times
\e^{-\epsilon a_-n^2+2\epsilon(m+1+a_--(1-a_-)u_0-a_-u+\half(v_1+v_2))n}.
\lnlabel\EQhmndef}
$$
The function $g(u,u_0;v_1,v_2)$ is the trace over the oscillators
$a_k$ and is given by
$$
\eqalignno{
g(u,u_0;v_1,v_2)
&=\chi^*c_\tV^2c_V^2c_Wg_{VV}(1)
g_{\tV\tV}(v_1-v_2)g_{VW_-}(u_0-u)g_{VW_-}(u_0-u+1)
\cr
&\quad\times
\prod_{j=1,2}g_{\tV V}(u-v_j)g_{\tV V}(u-v_j-1)g_{\tV W_-}(u_0-v_j).
\lnlabel\EQgdef}
$$

The sum
$$
h^{(i)}_m
=\sum_{n\in2\Z+m+i}K_n h_{mn}(u,u_0;v_1,v_2)
\eqlabel\EQhmdef
$$
can be calculated by the tricks described in Appendix~D of \Refs{\LPone}
and reads
$$
\eqalignno{
h^{(i)}_m={[0]_4\over2[u-u_0]_1}\e^{\epsilon k(u,u_0;v_1,v_2)}
&\biggl(\left[m+{v_1+v_2\over2}-u_0\right]'_4
\left\{{v_1+v_2\over2}-u\right\}_4
\cr
&\quad
+(-1)^i
\left[m+{v_1+v_2\over2}-u_0\right]'_1
\left\{{v_1+v_2\over2}-u\right\}_1
\biggr)
\lnlabel\EQhmexpr}
$$
with
$$
\{u\}_i=h_i(u;1)
$$
and
$$
k(u,u_0;v_1,v_2)
=\left(u-{v_1+v_2\over2}-2\right)^2+{r\over r-1}{v_1-v_2-2\over2}
-{r-1\over2r}-{(u-u_0)^2\over r}
+{1\over r-1}\left(u_0-{v_1+v_2\over2}-1\right)^2.
$$
From the identity
$$
\theta_4(u;\tau)\theta_3(v;\tau)\pm
\theta_3(u;\tau)\theta_4(v;\tau)
={2\theta_3(0;\tau)\theta_4(0;\tau)\over\theta^2_4(0;2\tau)}
\theta_{4,1}(u-v;2\tau)\theta_{4,1}(u+v;2\tau)
$$
we obtain
$$
\eqalignno{
{1\over C'}\left[v_2-v_1+1\over2\right]'_{4,1}
\left[m+{v_1+v_2-1\over2}\right]'_{4,1}
&=\theta_4\!\left({v_2+m\over 2r'};{\i\pi\over2\epsilon r'}\right)
\theta_3\!\left({v_1+m-1\over 2r'};{\i\pi\over2\epsilon r'}\right)
\cr
&\quad
\pm\theta_3\!\left({v_2+m\over 2r'};{\i\pi\over2\epsilon r'}\right)
\theta_4\!\left({v_1+m-1\over 2r'};{\i\pi\over2\epsilon r'}\right)
\lnlabel\EQhthetaid}
$$
with $C'$ given by (\EQCprimedef). By substituting (\EQhthetaid) into
(\EQhmexpr) and taking into account (\EQintertwiningtilde a) we obtain
that the form factor $F_{m-2\,m-1\,m}^{(i)}$ have the desired structure:
$$
\eqadvance\EQFmstruct
\eqalignno{
F_{m-2\,m-1\,m}^{(i)}(\theta_1,\theta_2)
&=(-)^{m-m_0}{c\over\chi}
h^{(i)}_m(u,u_0;v_1,v_2)g(u,u_0;v_1,v_2)
\cr
&=A^{(i)}_m(u,u_0;v_1,v_2)
\tdt_-(v_2-u_0+\texthalf)^m_{m-1}
\tdt_+(v_1-u_0+\texthalf)^{m-1}_{m-2}
\cr
&\quad
+B^{(i)}_m(u,u_0;v_1,v_2)
\tdt_+(v_2-u_0+\texthalf)^m_{m-1}
\tdt_-(v_1-u_0+\texthalf)^{m-1}_{m-2}.
\lnlabelno(\EQFmstruct a)}
$$
Here
$$
\eqalign{
A^{(i)}_m
&=G(u,u_0;v_1,v_2)((-1)^i a_1(u;v_1,v_2)+a_4(u;v_1,v_2)),
\cr
B^{(i)}_m
&=G(u,u_0;v_1,v_2)((-1)^i a_1(u;v_1,v_2)-a_4(u;v_1,v_2)),
\cr
G(u,u_0;v_1,v_2)
&=\e^{\epsilon k(u,u_0;v_1,v_2)}
{cC'[0]_4\over2\chi[u-u_0]_1}{g(u,u_0;v_1,v_2)
\over\tf(v_1-u_0+\half)\tf(v_2-u_0+\half)}
}\eqlabelno(\EQFmstruct b)
$$
with
$$
a_i(u;v_1,v_2)
={\{\half(v_1+v_2)-u\}_i\overtext[\half(v_2-v_1+1)]'_i}.
\eqlabelno(\EQFmstruct c)
$$
The factors $A^{(i)}_m$ and $B^{(i)}_m$ could be identified with
$F_{+-}(\theta_1,\theta_2)$ and $F_{-+}(\theta_1,\theta_2)$ if we could
prove that they are $u_0$ independent. To do it let us gather the
$u_0$ dependent factors of these expressions:
$$
e^{\epsilon k(u,u_0;v_1,v_2)}
{g_{VW_-}(u_0-u)g_{VW_-}(u_0-u+1)g_{\tV W_-}(u_0-v_1)g_{\tV W_-}(u_0-v_2)
\over[u-u_0]\tf(v_1-u_0+\half)\tf(v_2-u_0+\half)}.
$$
Note that the product
$$
g_{VW_-}(u_0-u)g_{VW_-}(u_0-u+1)
=-\e^{\epsilon((u-u_0)^2/r+u-u_0)}
{[u-u_0]\over(x^{2r};x^{2r})_\infty}
$$
in the numerator contains $[u-u_0]$ that cancels the same factor in the
denuminator. So, up to an exponential factor, the function
$\tf(v+\half)$ must be proportional to $g_{\tV W_-}(-v)$. This uniquely
determines the prefactor $\tf(u)$ in the form~(\EQintertwiningtilde b).

Finally, we obtain
$$
\eqadvance\EQtwoparticleff
\eqalign{
F^{(i)}_{+-}(\theta_1,\theta_2)
&=G(u;v_1,v_2)((-1)^i a_1(u;v_1,v_2)+a_4(u;v_1,v_2)),
\cr
F^{(i)}_{-+}(\theta_1,\theta_2)
&=G(u;v_1,v_2)((-1)^i a_1(u;v_1,v_2)-a_4(u;v_1,v_2)),
\cr
G(u;v_1,v_2)
&=cc_\tV^2c_V^2c_Wg_{VV}(1)[0]_4
{\chi^*\over2\chi}{(x^{2r-2};x^{2r-2})_\infty\over(x^{2r};x^{2r})_\infty}
\e^{\epsilon k'(u;v_1,v_2)}
\cr
&\quad\times
g_{\tV\tV}(v_1-v_2)\prod_{j=1,2}g_{\tV V}(u-v_j)g_{\tV V}(u-v_j-1)
}\eqlabelno(\EQtwoparticleff a)
$$
with
$$
k'(u;v_1,v_2)
=\left(u-{v_1+v_2\over2}-{3\over2}\right)^2
+{v_1-v_2-3/2\over2}
-{1\over r-1}\left(v_1-v_2-1\over2\right)^2
-{1\over2r(r-1)}.
\eqlabelno(\EQtwoparticleff b)
$$
The other two components, $F_{++}$ and $F_{--}$, of the form factor vanish
in consistency with the charge conservation modulo~4.

Now consider the limit $\theta_2\to\theta_1-\i\pi$ (or $v_2\to v_1-1$).
From (\EQevnorm) we obtain the equation
$$
F^{(i)}_{\pm\mp}(\theta,\theta'-\i\pi)
={2i\over\theta'-\theta}\langle\sigma^z\rangle^{(i)}+O(1),
$$
which is a particular case of the Smirnov form factor axioms\refs{\Smirnovbook}.
So, as a check of our construction we can calculate the expectation value
of the spin operator at a given link in a given ground state sector. This
reproduces the well-known Baxter--Kelland formula\refs{\BK}
$$
\langle\sigma^z\rangle^{(i)}
=(-)^i{(x^2;x^2)^2_\infty\over(-x^2;x^2)^2_\infty}
{(-x^{2r};x^{2r})^2_\infty\over(x^{2r};x^{2r})^2_\infty}.
\eqlabel\EQBaxterKelland
$$
In contrast to~\Refs{\LPone} this formula was obtained here without any
integrations.

\nsec Discussion

The main result of the paper is the bosonization recipe for the form
factors consisting of the intertwining algebra for vertex operators
(\EQToperators--\skipsec\EQvfvo), (\EQintertwiningtilde), (\EQDeltauzero)
and the bosonization prescription for the $\Lambda$ operator (\EQLambdafin),
(\EQLambdacoeffs).

Of course, this recipe is based on several conjectures. We conjectured
that the vertex operator algebras correctly describe the respective
models, and that the bosonization procedure provides relevant
representations of the vertex oprator algebra. Another problem is that
it seems to be very difficult to obtain more or less simple general
formulas for correlation functions and form factors, as it was done for
the six-vertex model\refs{\JMbook}. Moreover, the bosonization of the
eight-vertex operators depends on two additional parameters $m$ and $u_0$,
while the values of correlation functions and form factors are, of course,
$m$ and $u_0$ independent.%
\nfootnote{To be more precise, the parameter $u_0$ becomes a physical
parameter in a model with a boundary\refs{\Hara}.}

In the case of the six-vertex/XXZ model both vertex operator approach
and bosonization technique were based on the $U_q(\widehat{sl}_2)$
symmetry of the model\refs{\JMbook} which was, in turn, based on the
low-temperature expansions\refs{\DFJMN} and the theory of crystal
bases\refs{\KKMMNN}. The symmetry of the eight-vertex model was
conjectured to be the elliptic algebra
${\cal A}_{q,p}(\widehat{sl}_2)$\refs{\FIJKMY}, but neither low-temperature
expansion check was performed nor bosonization of this algebra was
acheived. The only encouraging result is its relation to the elliptic
algebra $E_{\tau,\eta}(\widehat{sl}_2)$ describing the SOS
model\refs{\Frons,\JKOS}. One might expect a natural solution of the problem
of $m$ and $u_0$ independence in this framework.

Another approach to calculation of correlation functions of lattice
models is based on the algebraic Bethe Ansatz method and seems to be
more rigorous\refs{\KMT,\IKMT}. Though generalization of this approach
to the eight-vertex model encounder significant difficulties, in
principle, it could provide a more direct way to obtaining correlation
functions and form factors.

An interesting feature of our construction is that two integrals, one
of the $X$ type and the other of the $\tX$ type, in the form factors
can be taken by use of the trick of Sec.~5. This feature shows
promise of obtaining more explicit formulas for correlation functions
and form factor. Note that the possibility of explicit taking two
integrals is, probably, a general feature of form factors\refs{\NPT},
though in the usual bosonization schemes it requires very sophisticated
calculations.

\nsec Acknowledgments

The author is grateful to A.~Belavin, S.~Lukyanov, A.~Odesskii, and
Ya.~Pugai for fruitful discussions and to M.~Jimbo and T.~Miwa for
their hospitality at RIMS, Kyoto University in July, 1999, where many
interesting discussions took place. The work was supported by INTAS
under the grant INTAS--OPEN--97--1312, by CRDF under the grant
RP1--2254 and by RFBR under the grants 99--01--01169 and 00--15--96579.

\secno=-1
\sec* Appendix~A. Consistency of the vertex operator algebra

Here we check consistency of the relations for the operators
$T(u_0)_{mn}$, $T(u_0)^{mn}$ with the algebras of vertex operators for
the eight-vertex and SOS models.

Consider the product
$$
T(u_0)_{mn'}\Phi(u_1)^{n'}_s\Phi(u_2)^s_n
\eqlabel\EQTPhiPhi
$$
and try to reverse the order of the operators. One can do it in two
ways. In the first way one can first push the operator $T(u_0)_{mn}$
through the vertex operators by use of (\EQvfvo), and then exchange the
vertex operators by use of (\EQevcommutation):
$$
\eqalignno{
(\EQTPhiPhi)
&=\sum_{\ve_1\ve_2}t_{\ve_1}(u_1-u_0)^{n'}_st_{\ve_2}(u_2-u_0)^s_n
\Phi_{\ve_1}(u_1)\Phi_{\ve_2}(u_2)T(u_0)_{mn}
\cr
&=\sum_{\ve_1\ve_2\ve'_1\ve'_2}
t_{\ve_1}(u_1-u_0)^{n'}_st_{\ve_2}(u_2-u_0)^s_n
R(u_1-u_2)^{\ve'_1\ve'_2}_{\ve_1\ve_2}
\Phi_{\ve'_2}(u_2)\Phi_{\ve'_1}(u_1)T(u_0)_{mn}.
\lnlabel\EQTPhiPhiIway}
$$
In the second way one can first commute vertex operators according to
(\EQsoscommutation) and then push $T(u_0)_{mn}$ by use of (\EQvfvo):
$$
\eqalignno{
(\EQTPhiPhi)
&=\sum_{s'}\W(s,n;s',n'|u_1-u_2)T(u_0)_{mn'}
\Phi(u_2)^{n'}_{s'}\Phi(u_1)^{s'}_n
\cr
&=\sum_{s'\ve_1\ve_2}
\W(s,n;s',n'|u_1-u_2)t_{\ve_2}(u_2-u_0)^{n'}_{s'}t_{\ve_1}(u_1-u_0)^{s'}_n
\Phi_{\ve'_2}(u_2)\Phi_{\ve'_1}(u_1)T(u_0)_{mn}.
\lnlabel\EQTPhiPhiIIway}
$$
The last lines of (\EQTPhiPhiIway) and (\EQTPhiPhiIIway) coincide because
of the vertex-face correspondence relation~(\EQvertexface). Consistency
with other commutation relations for vertex operators is checked in the
same way.

Now let us check consistency of the relation between the shift
operators (\EQvfctm) with commutation relations. We have
$$
\eqalignno{
\rho\Phi_\ve(u-2)
&=\sum_{n\in2\Z+m+i}T(u_0)_{mn}{\rho_{mn}\overtext[m]'}T(u_0)^{mn}\Phi_\ve(u-2)
\cr
&=\sum_{n\in2\Z+m+i}\sum_{n'=n\pm1}t_\ve(u-u_0-2)^n_{n'}
T(u_0)_{mn}{\rho_{mn}\overtext[m]'}\Phi(u-2)^n_{n'}T(u_0)^{mn'}
\cr
&=\sum_{n\in2\Z+m+i}\sum_{n'=n\pm1}{[n]\over[n']}t_\ve(u-u_0-2)^n_{n'}
T(u_0)_{mn}\Phi(u)^n_{n'}{\rho_{mn'}\overtext[m]'}T(u_0)^{mn'}
\cr
&=\sum_{n\in2\Z+m+i}\sum_{n'=n\pm1}t'_\ve(u-u_0)^n_{n'}
T(u_0)_{mn}\Phi(u)^n_{n'}{\rho_{mn'}\overtext[m]'}T(u_0)^{mn'}
\cr
&=\sum_{n'\in2\Z+m+1-i}\sum_{n=n'\pm1}\sum_{\ve'}
t'_\ve(u-u_0)^n_{n'}t^*_{\ve'}(u-u_0)^{n'}_n
\Phi_{\ve'}(u)T(u_0)_{mn'}{\rho_{mn'}\overtext[m]'}T(u_0)^{mn'}
\cr
&=\sum_{n'\in2\Z+m+1-i}
\Phi_\ve(u)T(u_0)_{mn'}{\rho_{mn'}\overtext[m]'}T(u_0)^{mn'}
\cr
&=\Phi_\ve(u)\rho.
\lnlabel\EQPhirhocheck}
$$
Here (\EQstarprimeexplicit) and (\EQstarprimedef) were used.
This check (in different notations) was one of the key results
of~\Refs{\LPone}.

The check for the $\Psi$ operators is slightly different because it
essentially uses the $m$ independence of $\rho$. Let $v_0=u_0+\Delta
u_0$. Then
$$
\eqalignno{
\Psi^*_\ve(u)\rho
&=\sum_{n\in2\Z+m+i}\Psi^*_\ve(u)T(u_0)_{mn}{\rho_{mn}\overtext[m]'}T(u_0)^{mn}
\cr
&=\sum_{n\in2\Z+m+i}\sum_{m'=m\pm1}\tdt^*_\ve(u-v_0)^m_{m'}
T(u_0)_{m'n}\Psi^*(u)^{m'}_m{\rho_{mn}\overtext[m]'}T(u_0)^{mn}
\cr
&=\sum_{n\in2\Z+m+i}\sum_{m'=m\pm1}
{[m']'\overtext[m]'}\tdt^*_\ve(u-v_0)^m_{m'}
T(u_0)_{m'n}{\rho_{m'n}\over[m']'}\Psi^*(u-2)^{m'}_mT(u_0)^{mn}
\cr
&=\sum_{n\in2\Z+m+i}\sum_{m'=m\pm1}\sum_{\ve'}
{[m']'\overtext[m]'}\tdt^*_\ve(u-v_0)^m_{m'}\tdt_{\ve'}(u-v_0-2)^{m'}_m
T(u_0)_{m'n}{\rho_{m'n}\over[m']'}T(u_0)^{mn}\Psi^*_{\ve'}(u-2)
\cr
&=\sum_{m'=m\pm1}\sum_{n\in2\Z+m'+1-i}\sum_{\ve'}
\tdt^*_\ve(u-v_0)^m_{m'}\tdt'_{\ve'}(u-v_0-2)^{m'}_m
T(u_0)_{m'n}{\rho_{m'n}\over[m']'}T(u_0)^{mn}\Psi^*_{\ve'}(u-2)
\cr
&=\sum_{m'=m\pm1}\sum_{\ve'}
\tdt^*_\ve(u-v_0)^m_{m'}\tdt'_{\ve'}(u-v_0-2)^{m'}_m
\rho\Psi^*_{\ve'}(u-2)
\cr
&=\rho\Psi^*_{\ve'}(u-2).
\lnlabel\EQPsirhocheck}
$$

It only remains to check consistency with the normalization conditions.
For example, let us multiplicate the second equation of~(\EQevnorm)
by~$T(u_0)^{mn'}$ from the left:
$$
\eqalignno{
T(u_0)^{mn}\Phi^{}_{\ve_1}(u)\Phi^*_{\ve_2}(u)
&=T(u_0)^{mn}\Phi_{\ve_1}(u)\Phi_{-\ve_2}(u-1)
\cr
&=\sum_{n',n''}t_{\ve_1}(u-u_0)^n_{n'}t_{-\ve_2}(u-u_0-1)^{n'}_{n''}
\Phi(u)^n_{n'}\Phi(u-1)^{n'}_{n''}T(u_0)^{mn''}
\cr
&=\sum_{n',n''}
(-1)^{n'-m_0}{n'-n''\over[n'']}
t_{\ve_1}(u-u_0)^n_{n'}t_{-\ve_2}(u-u_0-1)^{n'}_{n''}
\Phi(u)^n_{n'}\Phi^*(u)^{n'}_{n''}T(u_0)^{mn''}
\cr
&=\sum_{n'}t^{}_{\ve_1}(u-u_0)^n_{n'}t^*_{\ve_2}(u-u_0)^{n'}_n
\cr
&=\delta_{\ve_1\ve_2}.
\lnlabel\EQnormcheck}
$$
The normalization condition for $\Psi$s is checked similarly.

\secadvance
\sec* Appendix~B. Normal forms and commutation relations

Here some normal forms of operator products and some commutation
relations are listed for the operators $V$, $\bV$, $\tV$, $\tbV$, $W$.
Let $V_i(u)$ be some vertex operators of the form (\EQvosplit). Then
$$
V_i(u')V_j(u)
=z^{\prime\alpha_i\alpha_j}\lcolon V_i(u')V_j(u)\rcolon f_{ij}(u-u')
\eqlabel\EQnormViVj
$$
with
$$
\log f_{ij}(u)=\langle\phi^-_i(0)\phi^+_j(u)\rangle.
\eqlabel\EQnormfij
$$
For the operators with $A_k=-A_{-k}$ (like $V(u)$, $\bV(u)$, $\tV(u)$,
$\tbV(u)$, and $W(u)$ but not $W_\pm(u)$) we have
$$
f_{ij}(u)=f_{ji}(u).
\eqlabel\EQnormfijsym
$$
The commutation relations are given by
$$
V_i(u')V_j(u)=\left(z'\over z\right)^{\alpha_i\alpha_j}
{f_{ij}(u-u')\over f_{ji}(u'-u)}V_j(u)V_i(u'),
\qquad
(z=x^{2u},\ z'=x^{2u'}).
\eqlabel\EQcommutViVj
$$
We have
$$
\alpha_V=-\sqrt{r-1\over2r},
\qquad
\alpha_\tV=\sqrt{r\over2(r-1)},
\qquad
\alpha_W=\sqrt{2\over r(r-1)}
$$
and
$$
\eqalignno{
f_{VV}(u)
&=g(z;\epsilon,r),
\lnlabel\EQnormfVV
\cr
f_{V\bV}(u)
&=g^{-1}(xz)g^{-1}(x^{-1}z)
={(x^{2r-1}z;x^{2r})_\infty\over(xz;x^{2r})_\infty},
\lnlabel\EQnormfVbV
\cr
f_{\bV\bV}(u)
&=g(x^2z)g^2(z)g(x^{-2}z)
=(1-z){(x^2z;x^{2r})_\infty\over(x^{2r-2}z;x^{2r})_\infty},
\lnlabel\EQnormfbVbV
\cr
f_{\tV\tV}(u)
&=\tilde g(z)=(1-z)g^{-1}(z;\epsilon,r-1),
\lnlabel\EQnormftVtV
\cr
f_{\tV\tbV}(u)
&=\tilde g^{-1}(xz)\tilde g^{-1}(x^{-1}z)
={(x^{2r-1}z;x^{2r-2})_\infty\over(x^{-1}z;x^{2r-2})_\infty},
\lnlabel\EQnormftVtbV
\cr
f_{\tbV\tbV}(u)
&=\tilde g(x^2z)\tilde g^2(z)\tilde g(x^{-2}z)
=(1-z){(x^{-2}z;x^{2r-2})_\infty\over(x^{2r}z;x^{2r-2})_\infty},
\lnlabel\EQnormftbVtbV
\cr
f_{V\tV}(u)
&={(x^3z;x^4)_\infty\over(xz;x^4)_\infty},
\lnlabel\EQnormfVtV
\cr
f_{V\tbV}(u)
&=f_{\bV\tV}(u)=f^{-1}_{V\tV}(u+\texthalf)f^{-1}_{V\tV}(u-\texthalf)
=1-z,
\lnlabel\EQnormfVtbV
\cr
f_{\bV\tbV}(u)
&={1\over(1-x^{-2}z)(1-x^2z)},
\lnlabel\EQnormfbVtbV
\cr
f_{VW}(u)
&={\textstyle f_{V\tbV}(u+{r-1\over2})f_{V\bV}(u+{r\over2})}
={(x^{r-1}z;x^{2r})_\infty\over(x^{r+1}z;x^{2r})_\infty},
\lnlabel\EQnormfVW
\cr
f_{\bV W}(u)
&=f^{-1}_{VW}(u+\texthalf)f^{-1}_{VW}(u-\texthalf)
={(x^{r+2}z;x^{2r})_\infty\over(x^{r-2}z;x^{2r})_\infty},
\lnlabel\EQnormfbVW
\cr
f_{\tV W}(u)
&={\textstyle f_{\tV\tbV}(u+{r-1\over2})f_{\tV\bV}(u+{r\over2})}
={(x^rz;x^{2r-2})_\infty\over(x^{r-2}z;x^{2r-2})_\infty},
\lnlabel\EQnormftVW
\cr
f_{\tbV W}(u)
&=f^{-1}_{\tV W}(u+\texthalf)f^{-1}_{\tV W}(u-\texthalf)
={(x^{r-3}z;x^{2r-2})_\infty\over(x^{r+1}z;x^{2r-2})_\infty}
\lnlabel\EQnormftbVW}
$$
with $g(z)=g(z;\epsilon,r)$ from (\EQrhofactor).

Now list the most important commutation relations:
$$
\eqalignno{
V(u_1)V(u_2)
&=R_0(u_1-u_2)V(u_2)V(u_1),
\lnlabel\EQcommutVV
\cr
V(u_1)\bV(u_2)
&=-{[u_1-u_2+\half]\over[u_1-u_2-\half]}\bV(u_2)V(u_1),
\lnlabel\EQcommutVbV
\cr
\bV(u_1)\bV(u_2)
&={[u_1-u_2-1]\over[u_1-u_2+1]}\bV(u_2)\bV(u_1),
\lnlabel\EQcommutbVbV
\cr
\tV(u_1)\tV(u_2)
&=-R_0(u_2-u_1,\epsilon,r-1)\tV(u_2)\tV(u_1),
\lnlabel\EQcommuttVtV
\cr
\tV(u_1)\tbV(u_2)
&=-{[u_1-u_2-\half]'\overtext[u_1-u_2+\half]'}\tbV(u_2)\tV(u_1),
\lnlabel\EQcommuttVtbV
\cr
V(u_1)\tV(u_2)
&=-\i\tau(u_1-u_2)\tV(u_2)V(u_1),
\lnlabel\EQcommutVtV
\cr
V(u_1)\tbV(u_2)
&=-\tbV(u_2)V(u_1),
\lnlabel\EQcommutVtbV
\cr
\bV(u_1)\tV(u_2)
&=-\tV(u_2)\bV(u_1),
\lnlabel\EQcommutbVtV
\cr
\bV(u_1)\tbV(u_2)
&=\tbV(u_2)\bV(u_1),
\lnlabel\EQcommutbVtbV
\cr
\tbV(u_1)\tbV(u_2)
&={[u_1-u_2+1]'\overtext[u_1-u_2-1]'}\tbV(u_2)\tbV(u_1),
\lnlabel\EQcommuttbVtbV
\cr
V(u_1)W(u_2)
&={[u_1-u_2+{r+1\over2}]\over[u_1-u_2+{r-1\over2}]}W(u_2)V(u_1),
\lnlabel\EQcommutVW
\cr
\bV(u_1)W(u_2)
&={[u_1-u_2+{r-2\over2}]\over[u_1-u_2+{r+2\over2}]}W(u_2)\bV(u_1),
\lnlabel\EQcommutbVW
\cr
\tV(u_1)W(u_2)
&={[u_1-u_2+{r-2\over2}]'\overtext[u_1-u_2+{r\over2}]'}W(u_2)\tV(u_1),
\lnlabel\EQcommuttVW
\cr
\tbV(u_1)W(u_2)
&={[u_1-u_2+{r+1\over2}]'\overtext[u_1-u_2+{r-3\over2}]'}W(u_2)\tbV(u_1).
\lnlabel\EQcommuttbVW}
$$

\secadvance
\sec* Appendix~C. Proof of some identities for screening operators

Let us calculate the commutator $[\tY(u'),X(u)]$. Let
$\hat p=\sqrt{2r(r-1)}\P$. Consider the product
$$
\eqalignno{
\tY(u')X(u)
&={\epsilon^2\over\eta\eta'}\int_C{dv\over\i\pi}\int_{C'}{dv'\over\i\pi}\,
\tbV(v')\bV(v)
{[v'-u'+\half+\hat p]'\overtext[v'-u'-\half]'}
{[v-u+\half-\hat p]\over[v-u-\half]}
\cr
&={\epsilon^2\over4\eta\eta'}\int_C{dv\over\i\pi}\int_{C'}{dv'\over\i\pi}\,
\lcolon\tbV(v')\bV(v)\rcolon
{1\over\sh\epsilon(v'-v-\half)\sh\epsilon(v'-v+\half)}
\cr
&\quad\times
{[v'-u'+\half+\hat p]'\overtext[v'-u'-\half]'}
{[v-u+\half-\hat p]\over[v-u-\half]}.
\lnlabel\EQtYXprod}
$$
Here the contour $C$ goes along the imaginary axis to the left of the point
$u+\half$ and the contour $C'$ goes to the right of $u'+\half$ and to the
left of $v\pm\half$. The product $X(u)\tY(u')$ differs from (\EQtYXprod)
in that the contour for the variable $v'$ goes to the right of $v+\half$.
Hence, $[\tY(u'),X(u)]$ is the integral of the form (\EQtYXprod) but
with the contour for $v'$ enclosing poles $v\pm\half$. By taking the
integration in $v'$ by residues we obtain
$$
\eqalignno{
[\tY(u'),X(u)]
&={\epsilon\over2\eta\eta'\sh\epsilon}\int_C{dv\over\i\pi}
\biggl(\lcolon\tbV(v+\texthalf)\bV(v)\rcolon
{[v-u'-1+\hat p]'\overtext[v-u']'}
{[v-u+\half-\hat p]\over[v-u-\half]}
\cr
&\quad
-\lcolon\tbV(v-\texthalf)\bV(v)\rcolon
{[v-u'-2+\hat p]'\overtext[v-u'-1]'}
{[v-u+\half-\hat p]\over[v-u-\half]}\biggr)
\cr
&={\epsilon\over2\eta\eta'\sh\epsilon}\int_C{dv\over\i\pi}
\left(\textstyle
W(v+{r\over2})F(v+{r\over2},u,u';\hat p)
-W(v-{r\over2})F(v-{r\over2},u,u';\hat p)
\right).
\lnlabel\EQtYXcommint}
$$
Here $W(u)$ is the operator defined by (\EQWoperator) and
$$
F(v,u,u';\hat p)
={[v-u'-{r\over2}+1+\hat p]'\overtext[v-u'-{r\over2}]'}
{[v-u-{r-1\over2}-\hat p]\over[v-u-{r+1\over2}]}.
\eqlabel\EQFfunction
$$
The integrand in the last line of (\EQtYXcommint) is a total
difference. It means that
$$
[\tY(u'),X(u)]={\epsilon\over\eta\eta'\sh\epsilon}
\oint{dv\over2\pi\i}\,W(v)F(v;u,u',\hat p).
$$
There are two poles inside the contour of integration: $u'+{r\over2}$
and $u-{r-1\over2}$ and we finally obtain
$$
[\tY(u'),X(u)]
={\epsilon\over\eta\eta'\sh\epsilon}\biggl(
W_+(u')\,{[\hat p-1]'\overtext\partial[0]'}
{[u'-u+\half-\hat p]\over[u'-u-\half]}
+W_-(u)\,{[\hat p-1]\over\partial[0]}
{[u'-u+{3\over2}-\hat p]'\overtext[u'-u+\half]'}
\biggr)
\eqlabel\EQtYXcommut
$$
with $W_\pm(u)$ from (\EQYXcommutator b). Taking into account that
$[u+\hat p]=(-)^{m-n}[u+n]$ and $[u+\hat p]'=(-)^{m-n}[u+m]'$ on
$\F_{mn}$, we obtain (\EQYXcommutator a).

Consider the limit $u'\to u-\half$. According to (\EQtYXcommut) the
commutator has the pole at this point. This pole comes from pinching
the contour $C'$ in (\EQtYXprod) between $v-\half$ and $u'+\half$ for
$v$ being in the vicinity of the pole $u+\half$. Similarly, the pole at
$u'\to u+\half$ comes from pinching the contour $C'$ between $v+\half$
and $u'+\half$ for $v$ being in the vicinity of $u+\half$. For the
product $X(u)\tY(u')$ the contour for $v$ lies to the left of all poles
and the contour for $v'$ lies to the right of all poles. Without other
operators these contours can be moved far from all poles. It means that
the product $X(u)\tY(u')$ is always regular, while the poles of
(\EQtYXcommut) completely belong to the product $\tY(u')X(u)$. This
proves (\EQYXprodlim).

Let us prove (\EQlimformula). Consider the product
$$
\eqalignno{
\tY^k(u')X^l(u)
&=(-)^{\half k(k-1)+\half l(l-1)}
\left(\epsilon\over\eta\right)^k\left(\epsilon\over\eta'\right)^l
\int_{C^{\prime\times k}}{d^kw\over(i\pi)^k}
\int_{C^{\times l}}{d^lv\over(i\pi)^l}\,
\tbV(w_1)\ldots\tbV(w_k)\bV(v_l)\ldots\bV(v_1)
\cr
&\quad\times
\prod_{a=1}^k{[w_a-u'+\hat p+\half-2(k+1-a)]'\overtext[w_a-u'-\half]'}
\prod_{b=1}^l{[v_b-u-{3\over2}-\hat p+2b]\over[v_b-u-\half]}.
\lnlabel\EQtYkXldef}
$$
Because of the commutation relations (\EQcommutbVbV) and
(\EQcommuttbVtbV) the integrand can be symmetrized in all $w$'s and in
all $v$'s by use of the identity\refs{\JLMP}
$$
{1\over k!}\sum_{\sigma\in S_k}\prod_{i=1}^k[v_{\sigma(i)}-2i+2]
={[k]!\over k!\,[1]^k}\prod_{i<j}{[v_i-v_j]\over[v_i-v_j-1]}
\prod_{i=1}^k[v_i-k+1],
\qquad
[k]!=\prod_{i=1}^k[k].
\eqlabel\EQJMidentity
$$
We obtain
$$
\eqalignno{
&\tY^k(u')X^l(u)=
\cr
&\quad
=(-)^{\half k(k-1)+\half l(l-1)}
{[k]'!\,[l]!\overtext k!\,l!\,[1]^{\prime k}[1]^l}
\left(\epsilon\over\eta\right)^k\left(\epsilon\over\eta'\right)^l
\int_{C^{\prime\times k}}{d^kw\over(i\pi)^k}
\int_{C^{\times l}}{d^lv\over(i\pi)^l}\,
\tbV(w_1)\ldots\tbV(w_k)\bV(v_l)\ldots\bV(v_1)
\cr
&\qquad\times
\prod_{a=1}^k{[w_a-u'+\hat p-k-\half]'\overtext[w_a-u'-\half]'}
\prod_{b=1}^l{[v_b-u-\hat p+l-\half]\over[v_b-u-\half]}
\prod_{a<a'}{[w_a-w_{a'}]'\overtext[w_a-w_{a'}+1]'}
\prod_{b<b'}{[v_b-v_{b'}]\over[v_b-v_{b'}+1]}.
\cr
&\lnlabel\EQtYkXlsym}
$$
Let $u'\to u-\half$. If any of $w_a$ approaches $u'+\half$, the
contours $C$ become pinched between the poles $w_a+\half$ and
$u+\half$. It means that $C'$ can be deformed into a sum of two
contours $C'_l$ that goes to the left of $u'+\half$ and a small circle
$C'_0$ around $u'+\half$. Let us calculate the integral
$$
\eqalignno{
&\mskip -20mu
\int_{C^{\prime k-1}}{d^{k-1}w\over(\i\pi)^{k-1}}
\oint_{C'_0}{dw_k\over\i\pi}
\tbV(w_1)\ldots\tbV(w_{k-1})\tbV(w_k)
\prod_{a=1}^k{[w_a-u'+\hat p-k-\half]'\overtext[w_a-u'-\half]'}
\prod_{a<a'}^k{[w_a-w_{a'}]'\overtext[w_a-w_{a'}+1]'}
\cr
&=\int_{C^{\prime k-1}}{d^{k-1}w\over(\i\pi)^{k-1}}
\oint_{C'_0}{dw_k\over\i\pi}
\tbV(w_1)\ldots\tbV(w_{k-1})\tbV(w_k)
\cr
&\quad\times
\prod_{a=1}^{k-1}{[w_a-u'+\hat p-k-\half]'\overtext[w_a-u'-\half]'}
\prod_{a<a'}^{k-1}{[w_a-w_{a'}]'\overtext[w_a-w_{a'}+1]'}
{[w_k-u'+m-k-\half]'\overtext[w_k-u'-\half]'}
\prod_{a=1}^{k-1}{[w_k-w_a]'\overtext[w_k-w_a-1]'}
\cr
&=2\int_{C^{\prime k-1}_l}{d^{k-1}w\over(\i\pi)^{k-1}}
\tbV(w_1)\ldots\tbV(w_{k-1})\tbV(u'+\texthalf)
{[\hat p-k]'\overtext\partial[0]'}
\prod_{a=1}^{k-1}{[w_a-u'+\hat p-k-\half]'\overtext[w_a-u'+\half]'}
\prod_{a<a'}^{k-1}{[w_a-w_{a'}]'\overtext[w_a-w_{a'}+1]'}
\cr
&=(-)^{\half(k-1)(k-2)}
\tX^{k-1}(u')\tbV(u'+\texthalf)
\left(\eta'\over\epsilon\right)^{k-1}
{(k-1)!\,[1]^{\prime k-1}[\hat p-k]'\overtext[k-1]'!\,\partial[0]'}.
\lnlabel\EQtbVres}
$$
We substituted $C'$ to $C'_l$ in the fourth line
because the integrand no more has a pole at $u'+\half$. It means that one
may substitute
$$
\int_{C^{\prime k}}
\to\int_{C^{\prime k}_l}
+\sum_{a=1}^k\int_{C^{\prime a-1}_l\times C'_0\times C^{\prime a-1}_l}
\to\int_{C^{\prime k}_l}
+\>k\int_{C^{\prime k-1}_l}\int_{C'_0}
$$
in (\EQtYkXlsym).

Similarly, $C=C_r-C_0$ with $C_r$ going to the right of all poles and
$C_0$ being a small circle around $u+\half$. We obtain
$$
\eqalignno{
&\mskip -20mu
\int_{C^{l-1}}{d^{l-1}v\over(\i\pi)^{l-1}}\oint_{C_0}{dv_l\over\i\pi}\,
\bV(v_l)\bV(v_{l-1})\ldots\bV(v_1)
\prod_{b=1}^l{[v_b-u-\hat p+l-\half]\over[v_b-u-\half]}
\prod_{b<b'}^l{[v_b-v_{b'}]\over[v_b-v_{b'}+1]}
\cr
&=2\int_{C^{l-1}_r}{d^{l-1}v\over(\i\pi)^{l-1}}
\bV(u+\texthalf)\bV(v_{l-1})\ldots\bV(v_1){[l-\hat p]\over\partial[0]}
\prod_{b=1}^{l-1}{[v_b-u-n+l-\half]\over[v_b-u+\half]}
\prod_{b<b'}^{l-1}{[v_b-v_{b'}]\over[v_b-v_{b'}+1]}
\cr
&=(-)^{\half(l-1)(l-2)}
\bV(u+\texthalf)Y^{l-1}(u)
\left(\eta\over\epsilon\right)^{l-1}
{(l-1)!\,[1]^{l-1}[l-\hat p]\over[l-1]!\,\partial[0]}.
\lnlabel\EQbVres}
$$
The situation is the same as for the integral in~$w$'s.

By substitute (\EQtbVres) and (\EQbVres) into (\EQtYkXlsym) we obtain
$$
\eqalignno{
\tY^k(u')X^l(u)
&=\left(\tX^k(u'-1)+(-)^{k-1}\tX^{k-1}(u')\tbV(u'+\texthalf)
{\epsilon\over\eta'}
{[k]'[\hat p-k]'\overtext[1]^{\prime k}\,\partial[0]'}\right)
\cr
&\quad\times
\left(Y^l(u+1)+(-)^{l-1}\bV(u+\texthalf)Y^{l-1}(u)
{\epsilon\over\eta}
{[l][\hat p-l]\over[1]^l\,\partial[0]}\right)
\cr
&=\tX^k(u'-1)Y^l(u+1)
+(-)^{k-1}\tX^k(u'-1)\bV(u+\texthalf)Y^{l-1}(u)
{\epsilon\over\eta}
{[l][\hat p-l]\over[1]^l\,\partial[0]}
\cr
&\quad
+(-)^{l-1}\tX^{k-1}(u')\tbV(u'+\texthalf)Y^l(u+1)
{\epsilon\over\eta'}
{[k]'[\hat p-k]'\overtext[1]^{\prime k}\,\partial[0]'}
\cr
&\quad
+(-)^{k+l}\tX^{k-1}(u')\tbV(u'+\texthalf)\bV(u+\texthalf)Y^{l-1}(u)
{\epsilon^2\over\eta\eta'}
{[k]'[\hat p-k]'\overtext[1]^{\prime k}\,\partial[0]'}
{[l][\hat p-l]\over[1]^l\,\partial[0]}.
\lnlabel\EQtYkXlsum}
$$
The first three terms in the last expression are regular, because the
contour of integrations of $\tX$'s can be moved left and those of $Y$'s
right far away from poles. The last term contains the product
$$
\eqalignno{
\tbV(u'+\texthalf)\bV(u+\texthalf)
&=\lcolon\tbV(u'+\texthalf)\bV(u+\texthalf)\rcolon
{1\over4\sh\epsilon(u'-u-\half)\sh\epsilon(u'-u+\half)}
\cr
&=-{1\over u'-u+\half}{W_-(u)\over4\epsilon\sh\epsilon}+O(1)
\qquad\hbox{for $u'\to u-\texthalf$}.
\lnlabel\EQtbVbVexpansion}
$$
Substituting (\EQtbVbVexpansion) into (\EQtYkXlsum) and restricting
the result to $\F_{mn}$ we obtain~(\EQlimformula).

\secadvance
\sec* Appendix~D. Guidelines of the proof of (\EQLambdafin)

Eq.~(\EQLambdafin) can be checked as follows. Let us write
out the equations (\EQLambdavocommut b) for $\Delta u_0=-1/2$ explicitly:
$$
\eqadvance\EQPsiLambdaexplicit
\eqalignno{
\Psi^*(u)^{m-2k+1}_{m-2k}\Lambda(u_0)^{m-2k}_m
&=\Lambda(u_0)^{m-2k+1}_{m-1}\Psi^*(u)^{m-1}_m
{[u_0-u-m+k-\half]'[k-1]'\overtext [u_0-u-\half]'[m-1]'},
\cr
&\quad
+\Lambda(u_0)^{m-2k+1}_{m+1}\Psi^*(u)^{m+1}_m
{[u_0-u+k-\half]'[m-k+1]'\overtext [u_0-u-\half]'[m+1]'}
\lnlabelno(\EQPsiLambdaexplicit a)
\cr
\Psi^*(u)^{m-2k-1}_{m-2k}\Lambda(u_0)^{m-2k}_m
&=\Lambda(u_0)^{m-2k-1}_{m-1}\Psi^*(u)^{m-1}_m
{[u_0-u-k-\half]'[m-k-1]'\overtext [u_0-u-\half]'[m-1]'}
\cr
&\quad
+\Lambda(u_0)^{m-2k-1}_{m+1}\Psi^*(u)^{m+1}_m
{[u_0-u+m-k-\half]'[k+1]'\overtext [u_0-u-\half]'[m+1]'}
\lnlabelno(\EQPsiLambdaexplicit b)}
$$
Let us substitute (\EQbosvertex) and (\EQLambdafin) into
(\EQPsiLambdaexplicit), then pull the operator $Y^{l-1}(u_0)$ to the
right in each term and omit as a common factor. We obtain
$$
\eqalignno{
\tV(u)\tX^{k-1}(u_0-\texthalf)W_-(u_0)
&=\tX^{k-2}(u_0-\texthalf)W_-(u_0)\tY(u)\tV(u)
{[u_0-u-m+k-\half]'[k+1]'\overtext [u_0-u-\half]'[m]'}
\cr
&\quad
+\tX^{k-1}(u_0-\texthalf)W_-(u_0)\tV(u)
{[u_0-u+k-\half]'[m-k+1]'\overtext [u_0-u-\half]'[m]'},
\cr
\tY(u)\tV(u)\tX^{k-1}(u_0-\texthalf)W_-(u_0)
&=\tX^{k-1}(u_0-\texthalf)W_-(u_0)\tY(u)\tV(u)
{[u_0-u-k-\half]'[m-k-1]'\overtext [u_0-u-\half]'[m]'}
\cr
&\quad
+\tX^kW_-(u_0)\tV(u)
{[u_0-u+m-k-\half]'[k+1]'\overtext [u_0-u-\half]'[m]'}.
}
$$
It can be checked that all integration contours in $\tX$'s and $\tY$'s
in all terms can be deformed to the same curve. It means that we shall
prove (\EQPsiLambdaexplicit) if we check equality of the integrands
symmetrized in all integration variables.

To do it, let us substitute definitions of the screening operators
into these equations. Then we have to render the products
of the operators to any given order by use of commutation relations
(\EQcommuttVtbV), (\EQcommuttbVtbV), (\EQcommutbVW), and (\EQcommuttbVW).
For example, the first term in the r.~h.~s.\ of the first equation
is proportional to
$$
\eqalignno{
&{[u_0-u-m+k-\half]'[k-1]'\overtext [u_0-u-\half]'[m]'}
\int{d^{k-1}w\over(\i\pi)^{k-1}}\,
\tbV(w_{k-1})\ldots\tbV(w_2)\tbV(w_1)W_-(u_0)\tV(u)
\cr
&\quad\times
{[w_1-u_0+1]'\overtext[w_1-u_0-1]'}
{[w_1-u-\half+m]'\overtext[w_1-u-\half]'}
\prod_{a=2}^{k-1}{[w_a-u_0+m-1-2(a-2)]'\overtext[w_a-u_0+1]'}.
\lnlabel\EQexampleofint}
$$
The next step is to symmetrize the integrand. For symmetrization in
$w_2,\ldots,w_{k-1}$
the identity~(\EQJMidentity) can be applied. Symmetrization in the position
of the remaining variable $w_1$ produces the sum:
$$
\eqalignno{
(\EQexampleofint)
&={[u_0-u-m+k-\half]'[1]'\overtext [u_0-u-\half]'[m]'}
{[k-1]'!\overtext(k-1)!\,[1]^{k-1}}
\sum_{j=1}^{k-1}\biggl({[w_j-u_0-1]'\overtext[w_j-u_0+1]'}
{[w_j-u-\half+m]'\overtext[w_j-u-\half]'}
\cr
&\quad\times
\prod_{\scriptstyle a<a'\atop\scriptstyle a,a'\ne j}^{k-1}
{[w_a-w_{a'}]'\overtext[w_a-w_{a'}-1]'}
\prod_{\scriptstyle a=1\atop\scriptstyle a\ne j}^{k-1}
{[w_a-u_0+m-k]'\overtext[w_a-u_0+1]'}
\prod_{a=1}^{j-1}{[w_j-w_a-1]'\overtext[w_j-w_a+1]'}\biggr).
}
$$
The same should be done for every other term. By equating the integrands in
both sides, we obtain some equalities. To shorten the formulas let us
introduce notations $z_a=w_a-u-\half$, $z_0=u_0-u-\half$
and assign $z_{ab}=z_a-z_b$, $z_{a0}=z_a-z_0$. We have
$$
\eqalignno{
&\prod_{a<a'}^{k-1}{[z_{aa'}]'\overtext[z_{aa'}-1]'}
\prod_{a=1}^{k-1}{[z_{a0}+m-k]'\overtext[z_{a0}+1]'}
{[z_a+1]'\overtext[z_a]'}
\cr
&\quad
={[z_0-m+k]'[1]'\overtext[z_0+1]'[m]'}
\sum_{j=1}^{k-1}
\prod_{\scriptstyle a<a'\atop\scriptstyle a,a'\ne j}^{k-1}
{[z_{aa'}]'\overtext[z_{aa'}-1]'}
\prod_{\scriptstyle a=1\atop\scriptstyle a\ne j}^{k-1}
{[z_{a0}+m-k]'\overtext[z_{a0}+1]'}
\prod_{a=1}^{j-1}{[z_{ja}-1]'\overtext[z_{ja}+1]'}
\cr
&\qquad
+{[z_0+k]'[m-k+1]'\overtext[z_0+1]'[m]'}
\prod_{a<a'}^{k-1}{[z_{aa'}]'\overtext[z_{aa'}+1]'}
\prod_{a=1}^{k-1}{[z_{a0}+m-k+1]'\overtext[z_{a0}+1]'},
\cr
&{[m]'\overtext[k]'}\sum_{j=1}^k
{[z_j-2k+m]'\overtext[z_j]'}
\prod_{\scriptstyle a<a'\atop\scriptstyle a,a'\ne j}^k
{[z_{aa'}]'\overtext[z_{aa'}-1]'}
\prod_{\scriptstyle a=1\atop\scriptstyle a\ne j}^k
{[z_{a0}+m-k]'\overtext[z_{a0}+1]'}
{[z_a+1]'\overtext[z_a]'}
\prod_{a=j+1}^k{[z_{aj}-1]'\overtext[z_{aj}+1]'}
\cr
&\quad
={[z_0-k]'[m-k-1]'\overtext[k]'[z_0+1]'}
\sum_{j=1}^k
{[z_j+m]'\overtext[z_j]'}
\prod_{\scriptstyle a<a'\atop\scriptstyle a,a'\ne j}^k
{[z_{aa'}]'\overtext[z_{aa'}-1]'}
\prod_{\scriptstyle a=1\atop\scriptstyle a\ne j}^k
{[z_{a0}+m-k-1]'\overtext[z_{a0}+1]'}
\prod_{a=1}^{j-1}{[z_{aj}-1]'\overtext[z_{aj}+1]'}
\cr
&\qquad
+{[z_0+m-k]'[k+1]'\overtext[1]'[z_0+1]'}
\prod_{a=1}^k{[z_{a0}+m-k]'\overtext[z_{a0}+1]'}
\prod_{a<a'}^k{[z_{aa'}]'\overtext[z_{aa'}+1]'}.
}
$$
These identities are proven by comparing residues of poles by induction
along the guidelines of Appendix~C of~\Refs{\LPone}. The proof is
straightforward but very cumbersome and we omit it.

\secadvance
\sec* Appendix~E. Trace functions

Here we list the constants $c_i$ and functions $g_{ij}(u)$ defined by
(\EQcgdef) for the operators $V(u)$, $\tV(u)$ and $W_-$. We assume that
$z=x^{2u}$.
$$
\eqalignno{
c_V
&={(x^6;x^4,x^4,x^{2r})_\infty(x^{2r+6};x^4,x^4,x^{2r})_\infty
\over(x^8;x^4,x^4,x^{2r})_\infty(x^{2r+4};x^4,x^4,x^{2r})_\infty},
\lnlabel\EQtracecV
\cr
c_\tV
&={(x^4;x^4,x^4,x^{2r-2})_\infty(x^{2r+6};x^4,x^4,x^{2r-2})_\infty
\over(x^6;x^4,x^4,x^{2r-2})_\infty(x^{2r+4};x^4,x^4,x^{2r-2})_\infty},
\lnlabel\EQtracectV
\cr
c_W
&={(x^2;x^{2r-2})_\infty\over(x^2;x^{2r})_\infty},
\lnlabel\EQtracecW
\cr
g_{VV}(u)
&={(x^2z;x^4,x^4,x^{2r})_\infty(x^{2r+2}z;x^4,x^4,x^{2r})_\infty
(x^6z^{-1};x^4,x^4,x^{2r})_\infty(x^{2r+6}z^{-1};x^4,x^4,x^{2r})_\infty
\over(x^4z;x^4,x^4,x^{2r})_\infty(x^{2r}z;x^4,x^4,x^{2r})_\infty
(x^8z^{-1};x^4,x^4,x^{2r})_\infty(x^{2r+4}z^{-1};x^4,x^4,x^{2r})_\infty},
\lnlabel\EQtracegVV
\cr
g_{\tV\tV}(u)
&={(z;x^4,x^4,x^{2r-2})_\infty(x^{2r+2}z;x^4,x^4,x^{2r-2})_\infty
(x^4z^{-1};x^4,x^4,x^{2r-2})_\infty(x^{2r+6}z^{-1};x^4,x^4,x^{2r-2})_\infty
\over(x^2z;x^4,x^4,x^{2r-2})_\infty(x^{2r}z;x^4,x^4,x^{2r-2})_\infty
(x^6z^{-1};x^4,x^4,x^{2r-2})_\infty(x^{2r+4}z^{-1};x^4,x^4,x^{2r-2})_\infty},
\cr
&\lnlabel\EQtracegtVtV
\cr
g_{\tV V}(u)
&={(x^3z;x^4,x^4)_\infty(x^7z^{-1};x^4,x^4)_\infty
\over(xz;x^4,x^4)_\infty(x^5z^{-1};x^4,x^4)_\infty},
\lnlabel\EQtracegtVV
\cr
g_{VW_-}(u)
&={(z;x^4,x^{2r})_\infty(x^{2r+2}z^{-1};x^4,x^{2r})_\infty
\over(x^2z;x^4,x^{2r})_\infty(x^{2r+4}z^{-1};x^4,x^{2r})_\infty},
\lnlabel\EQtracegVWm
\cr
g_{\tV W_-}(u)
&={(xz;x^4,x^{2r-2})_\infty(x^{2r+3}z^{-1};x^4,x^{2r-2})_\infty
\over(x^{-1}z;x^4,x^{2r-2})_\infty(x^{2r+1}z^{-1};x^4,x^{2r-2})_\infty}.
\lnlabel\EQtracegtVWm}
$$
List also some important combinations
$$
\eqalignno{
g_{VV}(1)
&={(x^4;x^4,x^4,x^{2r})^2_\infty(x^{2r+4};x^4,x^4,x^{2r})^2_\infty
\over(x^6;x^4,x^4,x^{2r})^2_\infty(x^{2r+2};x^4,x^4,x^{2r})^2_\infty},
\lnlabel\EQtrgVVone
\cr
g_{\tV V}(u)g_{\tV V}(u-1)
&={1\over(x^{-1}z;x^4)_\infty(x^5z^{-1};x^4)_\infty}
=x^{\half(u-\half)(u-{5\over2})}
{(x^4,x^4)_\infty\over
\theta_1\!\left({u\over2}-{1\over4};{\i\pi\over2\epsilon}\right)},
\lnlabel\EQtrgtVVtVV
\cr
g_{VW_-}(u)g_{VW_-}(u+1)
&=x^{-u^2/r+u}{[u]\over(x^{2r};x^{2r})_\infty}.
\lnlabel\EQtrgVWmVWm}
$$

\bigskip\allowbreak\bigskip\immediate\closeout\rfile
  \vbox{\secfont\noindent References\bigskip}\nobreak
  \catcode`@=11\input refs.tmp\catcode`@=12\bigskip

\end